\documentclass[preprint,12pt]{iopart}

\expandafter\let\csname equation*\endcsname\relax
\expandafter\let\csname endequation*\endcsname\relax

\pdfoutput=1

\usepackage[T1]{fontenc}
\usepackage{graphicx}
\usepackage{caption}
\usepackage{subcaption}
\usepackage{color}
\usepackage[backref=none]{hyperref}
\hypersetup{
  colorlinks,
  citecolor=red,
  filecolor=black,
  linkcolor=blue,
  urlcolor=magenta
}
\usepackage{amsmath}
\usepackage{amssymb}
\usepackage{lineno}
\usepackage{cite}

\allowdisplaybreaks[1]

\newcommand{\SZ}[1][0]{S_{#1}}

\newcommand{\deltad}{\delta_\mathrm{D}}
\newcommand{\taur}{\tau_\mathrm{R}}
\newcommand{\taub}{\tau_\mathrm{B}}
\newcommand{\tret}{\tau_\circlearrowright}
\newcommand{\dd}[1]{\!\mathrm{d}#1\,}
\newcommand{\td}[1]{\frac{\mathrm{d}}{\mathrm{d}t}}
\newcommand{\floor}[1]{{\left\lfloor\hspace{-.25mm} #1 \hspace{-.25mm}
    \right\rfloor}}
\newcommand{\ceiling}[1]{{\lceil\hspace{-.25mm} #1 \hspace{-.25mm}
    \rceil}}

\begin{document}


\title{L\'evy walks on lattices as multi-state processes}

\author{Giampaolo Cristadoro\dag, Thomas Gilbert\ddag, Marco
  Lenci\dag\S\ and David P.~Sanders$\|$} 

\address{
  \dag
  Dipartimento di Matematica, Universit\`a di Bologna, 
  Piazza di Porta S. Donato 5, 40126 Bologna, Italy
}
\address{
  \ddag\
  Center for Nonlinear Phenomena and Complex Systems,
  Universit\'e Libre  de Bruxelles, C.~P.~231, Campus Plaine, B-1050
  Brussels, Belgium
}
\address{
  \S\
  Istituto Nazionale di Fisica Nucleare, Sezione di
  Bologna, Via Irnerio 46, 40126 Bologna, Italy
}
\address{
  $\|$\
  Departamento de F\'isica, Facultad de Ciencias, Universidad
  Nacional Aut\'onoma de M\'exico,  Ciudad Universitaria,
  04510 M\'exico D.F.,
  Mexico
}

\date{Version of \today}

\begin{abstract}
  Continuous-time random walks combining diffusive scattering and
  ballistic propagation on lattices model a class of L\'evy walks. The
  assumption that transitions in the scattering phase occur
  with exponentially-distributed waiting times leads to a description of
  the process in terms of multiple states,
  whose distributions evolve according to a set of delay differential
  equations, amenable to analytic treatment. We obtain an exact
  expression  of the  mean squared displacement associated with such
  processes and discuss the emergence of asymptotic scaling laws in
  regimes of diffusive and superdiffusive (subballistic)
  transport, 
  emphasizing, in the latter case, the effect of initial
  conditions on the transport coefficients.
  Of particular interest is
  the case of rare ballistic propagation, in which case a regime of
  superdiffusion may lurk underneath one of normal diffusion.  
\end{abstract}

\submitto{J. Stat. Mech. Theor. Exp.}

\ead{
  \mailto{giampaolo.cristadoro@unibo.it},
  \mailto{thomas.gilbert@ulb.ac.be}, 
  \mailto{marco.lenci@unibo.it},
  \mailto{dpsanders@ciencias.unam.mx}
} 



\section{Introduction}

Stochastic processes in which independent particles scatter randomly at
finite speed and may occasionally propagate over a long distance in
single bouts are known as L\'evy walks. Use of these models has become
ubiquitous in the study of complex diffusive processes
\cite{Shlesinger:1995uk, Klafter:1996PhysTod, Klages:2008p11988, 
  Denisov:2012PhRvE85, Zaburdaev:2014Levy}. They are particularly
relevant to situations such that the probability of a
long jump decays slowly with its length
\cite{Weiss:1983RandomWalks}. The scale-free superdiffusive motion of
L\'evy walkers \cite{Shlesinger:1993fw}
has been identified as an efficient foraging strategy
\cite{Viswanathan:2011physics}, spurring a
large interest in these models, including with regards to human
mobility patterns \cite{Brockmann:2006p8047, Raichlen:2014evidence}. 

A central quantity in this formalism is the distribution of free
path lengths, which gives the probability that a particle propagates over
a distance $x$ between two successive scattering events. Its
asymptotic scaling determines how the moments of the displacement
asymptotically scale with time. Assuming the probability of a
free path of length $x$  scales asymptotically as $x^{-\alpha-1}$,
one finds, in the so-called velocity picture of L\'evy walks, the
following scaling laws for the mean squared displacement after time
$t$  \cite{Geisel:1985p8023, Shlesinger:1985p18111, Klafter:1987Stochastic,
  Wang:1992PhysRevA.45.8407, Zumofen:1993p804}:    

\begin{equation}
  \langle r^2 \rangle_t \sim
  \begin{cases}
    t^2\,, & 0 < \alpha < 1\,, \\
    t^2/\log t\,, & \alpha = 1\,,\\
    t^{3 - \alpha}\,, & 1 < \alpha < 2\,,\\
    t \log t\,, & \alpha = 2\,,\\
    t\,, & \alpha > 2\,;
  \end{cases}
  \label{eq:scalings}
\end{equation}

\noindent see below for a precise definition of this quantity. A
scaling parameter value $0<\alpha \leq 2$ is such that the variance 
of the distribution of free paths diverges, in which case the process is
often called scale-free \cite{Brockmann:2006p8047}. Correspondingly,
the asymptotic divergence of the mean squared displacement gives rise
to anomalous transport in the form of superdiffusion.

In a recent paper \cite{Cristadoro:2014transport}, we considered
L\'evy walks on lattices and generalized the standard description of
the velocity picture of L\'evy walks, according to which a new jump
event takes place as soon as the previous one is completed, to include an
exponentially-distributed waiting time separating successive jumps. As
emphasised in reference \cite{Cristadoro:2014transport}, this
additional phase induces a differentiation between the states of
particles which are in the process of completing a jump and those that
are waiting to start a new one. We call the former states propagating
and the latter scattering. The distinction between these states leads
to a new theoretical framework of L\'evy walks in terms of multistate
processes \cite{Landman:1977PNAS430}, whereby the generalized master
equation approach to continuous time random walks
\cite{Montroll:1965p243,   Kenkre:1973em} translates into a set of
delay differential equations for the corresponding distributions. 

It is the purpose of this paper to show that a complete
characterization of the solutions of such multistate processes can be
obtained, which yields exact time-dependent analytic expressions of
their mean squared displacement. These expressions can, on the one hand,
be compared with the asymptotic solutions already reported in
\cite{Cristadoro:2014transport}, and, on the other, also prove
useful when the asymptotic regime is not reached, which
is often the case with studies dealing with observational data.

Such a situation arises when the probability of a transition
from scattering to propagating states is small. A particle
will then spend most of its time undergoing transitions among 
scattering states, performing a seemingly standard continuous-time
random walk, only seldom undergoing a transition to a propagating
state, during which it moves ballistically over a distance distributed
according to the scaling parameter $\alpha$. If 
$\alpha$ is small enough ($\alpha \leq 2$), these occurrences,
although rare, have 
a dramatic effect on the asymptotic scaling properties of the mean
squared displacement, such that a crossover from normal to anomalous
diffusive transport is observed. This crossover time may in some 
situations, however, be much larger than the times accessible to
measurements. Here we provide 
analytic results which give a precise characterisation of the
emergence of an anomalous contribution out of a normally scaling
one, showing how the asymptotic scaling \eqref{eq:scalings} comes
about. This is of particular relevance to the regime $\alpha = 2$, for
which the logarithmic divergence in time of the mean squared
displacement is indeed very slow. Our results are tested and confirmed
by numerical simulations of the processes under consideration.

The paper is organized as follows. In \sref{sec:kproc}, we introduce
the multi-state description of L\'evy walks with respect to scattering
and propagating states, and define the transition probabilities
between them in terms of two parameters, one relating to the
probability of a transition from a scattering to a propagating state
and the other characterizing the asymptotic scaling of the free path
distribution. The fraction of particles in a scattering state evolves
according to a delay differential equation which is derived and solved
in \sref{sec:S0}. These results are exploited in \sref{sec:msd} to
obtain the mean squared displacement of the processes. A comparison
with numerical simulations is provided in
\sref{sec:numn2}. Conclusions are drawn in \sref{sec:con}.

\section{L\'evy walks as multi-state processes} 
\label{sec:kproc}

We consider a continuous-time random walk on a square lattice which
generalizes the standard model of Montroll \& Weiss
\cite{Montroll:1965p243} in that it assumes displacements to
non-neighbouring sites occur in a time span given by the ratio of the distance
traveled to the walker's speed $v$, which itself remains
fixed throughout. In this sense, the model is similar to the so-called
velocity picture of L\'evy walks \cite{Klafter:1995p686}, the
difference being that successive propagations, irrespective of their
lengths, are separated by random waiting times, which we assume 
have exponential distributions. The model itself is not new and, in
some sense, is a simplification of other models of L\'evy walks
interrupted by rests; see reference \cite{klafter:2011first}. As we
show below, however, the combination of a discretized spatial structure and 
exponentially-distributed waiting times yields a novel description
of the process in terms of delay differential equations amenable to
analytic treatment.

A natural distinction arises between the states of walkers which are
moving across the lattice structure and those at rest. We call
\emph{propagating} the state of a particle which is undergoing a
displacement phase and \emph{scattering} the
state of a particle at rest, waiting to start a new displacement. In
the framework of intermittent random walks
\cite{Benichou:2011intermittent}, the former state is usually referred
to as ballistic and the latter as diffusive or reactive, depending on
context. A scattering state may therefore be thought of as one 
associated with a local diffusive process bound to the scale $\ell$ of
the distance between neighbouring lattice sites. 

The interplay between the two states is as follows. A
particle in a propagating state switches to a scattering state upon
completing a displacement. A particle in a scattering state,  on the
other hand, can make transitions to both scattering and propagating
states; as soon as its randomly-distributed waiting time has elapsed,
it moves on to a neighbouring site and, in doing so, may switch to a
propagating state and continue its motion to the next site, or start
anew in a scattering state. 

Consider a $d$-dimensional cubic lattice of individual cells
$\mathbf{n}\in(\ell\, \mathbb{Z})^d$. The state of a walker at
position $\mathbf{n} \equiv (n_1,\dots,n_d)\ell$ and time $t$ can
take on a countable number of 
different values, $(k, j)$, specified by an integer $k\in\mathbb{N}$, and
direction $j \in \{1,\dots,z\}$, where $z \equiv 2d$ denotes the
coordination number of the  lattice, that is the number of different
lattice directions. States $(0, j)$ are associated with a
scattering state, irrespective of direction $j$, while states $(k, j)$
with $k \geq 1$ refer to propagating states, with $k$ being the
remaining number of lattice sites the walker will travel in direction
$j$ to complete its displacement.

Time evolution proceeds in steps, characterized by a
waiting-time density function and a transition probability. A particle
at position $\mathbf{n}$ in a scattering state will wait for a random
time $t$, exponentially distributed with mean\footnote{The subscript R
  in $\taur$ stands for residence as in ``residence time.'' In
  contrast, B in $\taub$ stands for ballistic as in ``ballistic
  propagation time.''} $\taur$, before updating 
its state to $(k,j)$ with probability $\rho_k/z$, simultaneously
changing its location to $\mathbf{n} + \mathbf{e}_j$, where
$\mathbf{e}_j$ denotes the lattice vector (of length $\ell$)
associated with direction $j$. In contrast, a particle in a
propagating state $(k, j)$, $k \geq 1$, will change its 
state to $(k - 1, j)$ after a time $\taub \equiv \ell/v$,
simultaneously moving from site $\mathbf{n}$ to $\mathbf{n} +
\mathbf{e}_j$. The waiting times associated with scattering
states are drawn from a standard Poisson process, while the renewal
process generated by the combination of scattering and propagating
states has arbitrary holding times, whose distribution is determined by
the transition probabilities $\rho_k$. 

The waiting-time density of the process is thus the function

\begin{equation}
  \psi_k(t) = 
  \begin{cases}
    \taur^{-1} e^{- t/\taur}, & k =0,\\
    \deltad(t -\taub), & k \neq 0,
  \end{cases}
  \label{eq:psik}
\end{equation}

\noindent where $\deltad(.)$ denotes the Dirac delta function. When a
step takes place, the transition probability to go from 
state $(k,j)$ to state $(k',j')$ is

\begin{equation}
  \mathsf{p}_{(k,j),(k',j')} = 
  \begin{cases}
    z^{-1} \rho_{k'},&  k =0,\\
    \delta_{k-1,k'} \delta_{j,j'} & k \neq 0,
  \end{cases}
  \label{eq:Pkj}
\end{equation}

\noindent where $\delta_{\cdot,\cdot}$ is the Kronecker symbol.

A particle which makes a transition from the scattering state to a
propagating state $(k,j)$ will therefore travel a distance $(k + 1) \ell$ away in 
direction $j$, until it eventually comes back to the scattering state
and may change directions at the next transition.

We now introduce a characterization of the transition
probabilities $\rho_k$ in terms of two parameters. The first,
which we refer to as the \emph{scattering parameter}, is denoted 
by $\epsilon$, $0 \leq \epsilon \leq 1$, and gives the total
probability of a transition from the scattering state, $k=0$, into a
propagating state, $k\geq1$:

\begin{equation}
  \sum_{k = 1}^\infty \rho_k = \epsilon,
  \label{eq:sumnuk}
\end{equation}

\noindent The remaining transition probability,

\begin{equation}  
  \rho_0 \equiv 1 - \epsilon,
  \label{eq:rho0}
\end{equation}

\noindent  is that of a transition from a scattering state into
another scattering state. The value $\epsilon = 0$ thus corresponds to
the absence of 
propagating states: the process is then a simple continuous-time
random walk with transitions to nearest neighbouring sites only and
exhibits normal diffusion with coefficient $\ell^2/(z \taur)$. The
opposite extreme, $\epsilon = 1$, assigns zero probability to
transitions from scattering to scattering states, which means that
every transition involves a displacement over a distance of at least
two sites. Scattering states remain populated, however, due to the decay
of propagating states. 

The second parameter, $\alpha >0$, is the \emph{scaling
  parameter} of the transition probabilities $\rho_k$, which controls
their asymptotic behaviour,
\begin{equation}
  \rho_k \propto k^{-\alpha - 1} \qquad (k\gg1), 
  \label{eq:rhokscaling}
\end{equation}
and determines the scaling law of the mean squared displacement
\eqref{eq:scalings}. The specific form of 
$\rho_k$ has no effect on this scaling law, but is does affect 
the time-dependent properties of the mean squared displacement. 
To be specific, we consider in this paper the following 
double-telescopic form for the
transition probabilities $\rho_k$:

\begin{equation}
  \rho_k = \epsilon (1 - 2^{1-\alpha})^{-1}
  \big[ k^{1-\alpha} - 2 (k + 1)^{1-\alpha} + (k + 2)^{1-\alpha}\big]
  \qquad (k\geq1);
  \label{eq:rhok2}
\end{equation}

\noindent its structure is particularly helpful for some of the
computations presented below and is motivated by our
study of anomalous transport in the infinite-horizon periodic Lorentz
gas \cite{Cristadoro:2014Machta}. In this respect, the model is
slightly different from that presented in reference
\cite{Cristadoro:2014transport}, where a 
simple telescopic structure of the transition probabilities was
used. 

For future reference, we define

\begin{equation}
  \label{eq:nuk}
  \nu_k = \sum_{j = k}^{\infty} \rho_{j},
\end{equation}

\noindent 
to be the probability of a transition to a state larger than or equal
to $k$, such that, in particular, $\nu_0 = 1$ and $\nu_1 =
\epsilon$, and, for the choice of transition probabilities
\eqref{eq:rhok2},

\begin{equation}
  \label{eq:nuk2}
  \nu_k = 
  \epsilon (1 - 2^{1-\alpha})^{-1}
  \big[ k^{1-\alpha} - (k + 1)^{1-\alpha} \big].
\end{equation}

\noindent We also note the following two identities: 

\begin{equation}
  \label{eq:rhok2id}
  \begin{split}
    \sum_{j = 1}^k \rho_j &=
    \epsilon \big \{1 - (1 - 2^{1-\alpha})^{-1} [(k + 1)^{1-\alpha} - (k +
    2)^{1-\alpha}] \big \},
    \\
    \sum_{j = 1}^k j \rho_j &=
    \epsilon (1 - 2^{1-\alpha})^{-1} [1 - (k + 1)^{2-\alpha} + (k +
    2)^{2-\alpha} - 2 (k+2)^{1 - \alpha}].
  \end{split}
\end{equation}

\section{Fraction of particles in the scattering state}
\label{sec:S0}

A quantity which plays a central role in the analysis of the process
generated by the waiting-time density \eqref{eq:psik} and transition
probabilities \eqref{eq:Pkj} is the average return time to the
scattering state, 

\begin{align}
  \tret &= \sum_{k = 0}^\infty \rho_k ( \taur + k \taub),
      \label{eq:tauret}
\end{align}

\noindent  
The average return time is finite only
when\footnote{Generally speaking, $\rho_k$ must decay asymptotically
  faster than $k^{-2}$ for the average return time to be finite.}
$\alpha > 1$. The process is then said to be \emph{positive recurrent}. In
the remainder, we restrict our attention to 
this range of parameter values. The null-recurrent case, when
$0<\alpha\leq1$, for which the average return time to the scattering
state diverges, is considered in reference \cite{Cristadoro:2014transport}. 

The occupation probability of particles at site $\mathbf{n}$ and
time $t$, $P(\mathbf{n}, t)$, is a sum of the probabilities over the
different states, $P_{k,j}(\mathbf{n}, t)$:

\begin{equation}
  P(\mathbf{n}, t)  = \sum_{j=1}^{z}
  \sum_{k=0}^\infty P_{k,j}(\mathbf{n}, t).
  \label{eq:Pn}
\end{equation}

\noindent In reference \cite{Cristadoro:2014transport}, we obtained
the following set of delay differential equations for the
time-evolution of these occupation probabilities:

\begin{align}
  \partial_t P_{0,j}(\mathbf{n}, t)  
  & =  
    \frac{1}{z \taur} 
    \sum_{j' = 1}^{z}
    \sum_{k=0}^\infty   \rho_{k} 
    P_{0, j'}(\mathbf{n} - (k+1) \mathbf{e}_j, t - k \taub)    
    - \frac{1}{\taur}P_{0, j}(\mathbf{n}, t)
    \cr
  & \quad + 
    \sum_{k = 0}^\infty 
    \sigma_{k,j}(\mathbf{n} - k \mathbf{e}_{j}, t  - k \taub )
    \,,
    \label{eq:dtP0}
  \\
  \partial_t P_{k,j}(\mathbf{n}, t)  
  & =  
    \frac{1}{z \taur}
    \sum_{j' = 1}^{z}
    \sum_{k' = 1}^\infty   \rho_{k + k'-1} 
    \Big[
    P_{0,j'}(\mathbf{n} - k' \mathbf{e}_{j}, t - (k'-1) \taub)
    \cr
  & \quad  
    - P_{0,j'}(\mathbf{n} - k' \mathbf{e}_{j}, t - k' \taub)
    \Big]
    + 
    \sum_{k' = 0}^\infty 
    \Big[
    \sigma_{k+k'\!, j}(\mathbf{n} - k' \mathbf{e}_{j}, t  - k' \taub )  
    \cr
  & \quad -
    \sigma_{k+k'\!, j}(\mathbf{n}  - k' \mathbf{e}_{j}, t - (k' +1)
    \taub)   
    \Big]
    \,,
    \label{eq:dtPkj}
\end{align}

\noindent where the inclusion of terms $\sigma_{k,j}$ accounts for the
possibility of external sources, such that
$\sigma_{k,j}(\mathbf{n}, t)  \geq 0$ is the rate of injection of
particles at site $\mathbf{n}$ and time $t$ in the state $(k,j)$.

To simplify matters, we assume that these source terms 
are independent of the lattice direction and write
$\sigma_{k,j}(\mathbf{n}, t) \equiv z^{-1} \sigma_{k}(\mathbf{n},
t)$. Furthermore, we will usually let $\sigma_{k,j}(\mathbf{n}, t) =
0$ for $k\geq 1$, which amounts to assuming that particles are 
injected only in a scattering state. The injection of particles in a
propagating state is easily treatable as well, and will be considered
explicitly in order to initiate the process in a stationary
state of equations \eqref{eq:dtP0}--\eqref{eq:dtPkj}.
We will,
however, limit such considerations to this specific choice and
avoid the possibility that source terms interfere with the asymptotic
scaling of  the process generated by particles initially injected in a
scattering state, as might arise from alternative choices of injection
rates of propagating states. 

Of particular interest is the fraction of particles in
the scattering state,

\begin{equation}
  \SZ(t) \equiv   
  \sum_{\mathbf{n}\in\mathbb{Z}^d}   \sum_{j = 1}^{z} P_{0, j}(\mathbf{n}, t),
  \label{eq:S0} 
\end{equation}

\noindent whose time-evolution is described by the following delay differential
equation:

\begin{equation}
  \dot \SZ(t) 
  = \taur^{-1} \sum_{k=1}^\infty \rho_k \SZ(t - k \taub) - 
  \epsilon \taur^{-1} \SZ(t)
  + \sum_{k = 0}^{\infty} \sigma_k(t - k \taub).
  \label{eq:dtS0}
\end{equation}

\noindent Here, $\sigma_k(t - k \taub)$ denotes a source term for the
rate of injection of particles in state $k$, irrespective of their
positions, that is, $\sigma_k(t - k \taub) = \sum_{\mathbf{n}}
\sigma_k(\mathbf{n}, t - k \taub)$.
The first two terms on the
right-hand side of this equation have the typical gain and loss
structure of jump processes. Indeed, the second term corresponds to
particles lost by scattering states due to transitions to 
propagating states, which occur at rate $\epsilon/\taur$. Each such
transition is gained back after a delay given by the length of the
ballistic segment in a propagating state. The sum of those terms yields the first
term on the right-hand side of equation \eqref{eq:dtS0}.

Before turning to the integration of this differential equation, we
note that the asymptotic value of $\SZ(t)$, $t\to\infty$, has a simple
expression. By ergodicity of the process, the equilibrium ratio of
particles in the scattering state is given by the ratio
between the average time spent in the scattering state, $\taur$, and
the average return time to it, $\tret$, equation \eqref{eq:tauret}, 

\begin{equation}
  \lim_{t\to\infty} \SZ(t) = \frac{\taur}{\tret}.
  \label{eq:S0largetime}
\end{equation}

\noindent Assuming a positive recurrent process, such as when the
transition probabilities $\rho_k$ are specified by equation
\eqref{eq:rhok2} with $\alpha>1$, the average return time 
$\tret = \taur + \epsilon (1 - 2^{1-\alpha})^{-1} \taub$
is finite, so 
that the asymptotic fraction of particles in a scattering state is
strictly positive. For the particular choice of the parameters
$\rho_k$, equation \eqref{eq:rhok2}, the return time is given
by equation \eqref{eq:tauret} and equation \eqref{eq:S0largetime}
becomes

\begin{equation}
  \lim_{t\to\infty} \SZ(t) = \frac{\taur}
  {\taur + \epsilon (1 - 2^{1-\alpha})^{-1} \taub}
  \, .
  \label{eq:S0rhok2}
\end{equation}

\noindent Note, however, that the convergence to this
asymptotic value follows a power law whose exponent goes to
zero as $\alpha\to 1$; see \fref{fig:Szexact}. Furthermore, $\SZ(t)$
converges to $0$ when the scaling parameter falls into the null 
recurrent regime, $0< \alpha \leq 1$.  

Correspondingly, the fraction of particles in 
the propagating states,

\begin{equation}
  \SZ[k](t) \equiv   
  \sum_{\mathbf{n}\in\mathbb{Z}^d}   \sum_{j = 1}^{z} P_{k,
    j}(\mathbf{n}, t),
  \label{eq:Sk} 
\end{equation}

\noindent 
tends to

\begin{equation}
  \lim_{t\to\infty} \SZ[k](t) = \frac{\nu_k \taub}
  {\taur + \epsilon (1 - 2^{1-\alpha})^{-1} \taub}
  \, ,
  \label{eq:Skrhok2}
\end{equation}

\noindent
which follows from the observation that a particle that
makes a transition to state $k$ spends an equal amount of time in all
states $j$ such that $1\leq  j \leq k$.

\subsection{Time-dependent fraction of particles in the scattering
  state \label{sec:S0t} 
}

The integration of  equation \eqref{eq:dtS0} relies on the
specification of initial conditions. For instance, the choice\footnote{The 
  discussion below can be easily generalized, by   linear
  superposition, to processes where particles in a scattering 
  state are continuously injected into the process.}

\begin{equation}
  \label{eq:S0ic}
  \SZ(t) = 
  \begin{cases}
    0, & t < 0, \\
    1, & t = 0,
  \end{cases}
\end{equation}

\noindent corresponds, when all particles are injected at the lattice
origin, to the injection rate  

\begin{equation}
  \label{eq:sigma0}
  \sigma_k(\mathbf{n}, t) =
  \deltad(t) \delta_{k,0} \delta_{\mathbf{n}, \mathbf{0}},
\end{equation}

\noindent
and will be henceforth referred to as the \emph{all-scattering initial condition}.

The time-dependent fraction of particles in the scattering state,
$\SZ(t)$, $t>0$, may then be obtained by the method of steps
\cite{Driver:1977ordinary}, which consists of 
integrating the differential equation \eqref{eq:dtS0} successively
over the intervals $k \taub \leq t \leq 
(k+1) \taub$, $k \in \mathbb{N}$, matching the solutions at the upper
and lower endpoints of successive intervals. 
For the all-scattering initial condition \eqref{eq:S0ic}, the fraction
of particles in the scattering state is, for times $t\geq 0$, 

\begin{equation}
  \SZ(t) = e^{-\epsilon  t/\taur} + \sum_{k=1}^{\floor{t/\taub}} 
  e^{-\epsilon (t-k \taub)/\taur}  \sum_{n = 1}^{k} a_{(n|k)} 
  \taur^{-n} (t - k \taub)^n,
  \label{eq:solS0}
\end{equation}

\noindent where $\floor{t/\taub}$ (resp. $\ceiling{t/\taub}$, used below)
denotes the largest (resp. smallest) integer smaller 
(resp. larger) than or equal to ${t/\taub}$, and each coefficient $a_{(n|k)}$ is
the sum of all possible combinations of products $\rho_{i_1} \dots
\rho_{i_n}$, where the sequences $\{i_j\}_{j=1}^n$  are such that $i_1
+ \dots + i_n = k$, 

\begin{equation}
  a_{(n|k)} = \frac{1}{n!}
  \sum_{
    \substack{
      1 \leq i_1,\dots,i_n \leq k
      \\
      i_1 + \dots + i_n = k
    }}
  \prod_{j = 1}^n \rho_{i_j} .
  \label{eq:akndef}
\end{equation}

\begin{figure}[h]
  \centering
  \begin{subfigure}[b]{0.49\textwidth}
    \includegraphics[width=\textwidth]
    {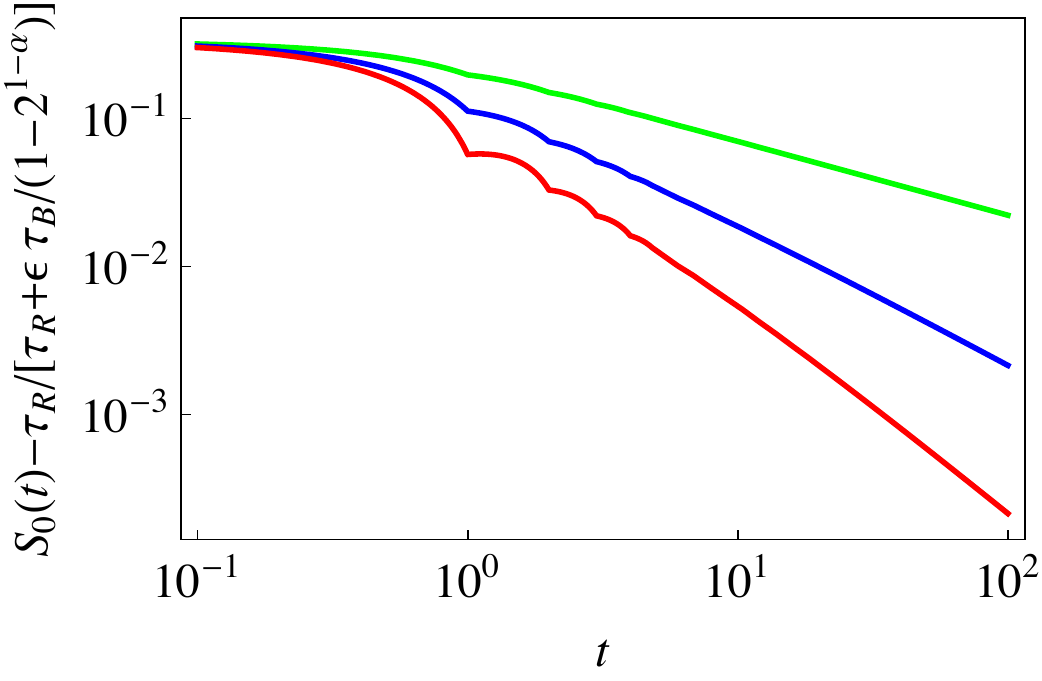}
    \caption{
      $\epsilon(1 - 2^{1-\alpha})^{-1} = 1$} 
    \label{fig:Szexact}
  \end{subfigure}
  \begin{subfigure}[b]{0.49\textwidth}
    \includegraphics[width=\textwidth]
    {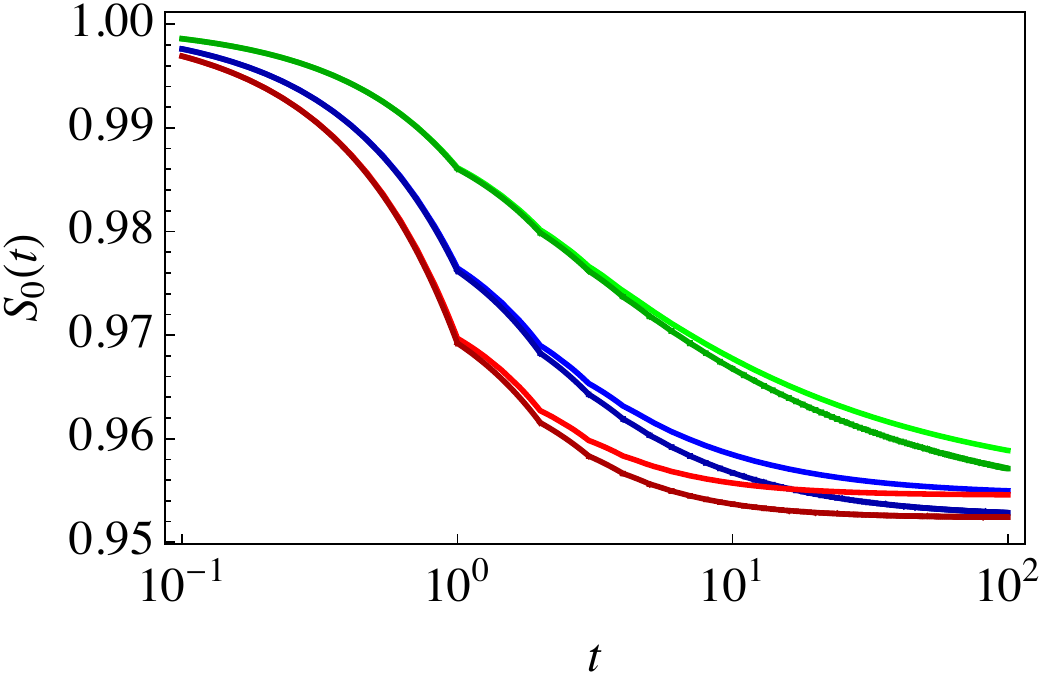}
    \caption{
      $\epsilon(1 - 2^{1-\alpha})^{-1} = 5 \times 10^{-2}$}
    \label{fig:Szapprox}
  \end{subfigure}
  \caption{
    Convergence in time of the fraction of
    particles in a scattering state, $\SZ(t)$, computed from equation 
    \eqref{eq:solS0}, to the asymptotic value \eqref{eq:S0rhok2}. The
    transition probabilities are taken according to equation
    \eqref{eq:rhok2}, with $\taub = 1$ and $\taur = 1 + \epsilon(1 -
    2^{1 - \alpha})^{-1}$. The value of the scattering parameter is
    chosen such that $\epsilon/(1 - 2^{1-\alpha}) = 1$ (left panel)
    and $5 \times 10^{-2}$ (right panel). The left panel shows the
    difference between $\SZ(t)$ and its asymptotic value. In both
    panels, the scaling parameter takes the values $\alpha = 3/2$ (green 
    curve), $2$ (cyan curve), and $5/2$ (red curve). 
    The right panel corresponds to a perturbative regime, such that
    $\SZ(t)$ can be approximated by a first-order polynomial in
    $\epsilon$ (darker curves) whose first-order coefficient varies with
    time; see equation~\eqref{eq:S0expepsi}. The agreement between the
    exact and the approximate solutions improves as the
    value of $\epsilon$ decreases.
  }   
  \label{fig:Sz}
\end{figure}

\noindent  The contributions to
$a_{(n|k)}$ consist of all distinct ways of travelling a
distance $k$ in $n$ steps, divided by their number of
permutations; the derivation of equation \eqref{eq:solS0} is provided
in \ref{app:1}. 
A comparison between the time-dependent fraction of particles in the
scattering state \eqref{eq:solS0} and the asymptotic value
\eqref{eq:S0largetime} is illustrated in \fref{fig:Szexact}, where the
difference between the two expressions is plotted vs.~time for
different values of the scaling parameter $\alpha$ and a specific
choice of the scattering parameter $\epsilon$, with the transition
probabilities $\rho_k$ specified by equation \eqref{eq:rhok2}. 

\subsection{Small-parameter expansion \label{sec:S0exp}}

Suppose that the scattering parameter is small, $\epsilon \ll 1$, so
that the overwhelming majority of transitions occur between scattering
states, and only rarely does a particle make an excursion into a
propagating state, which may, however, last long, depending on the value
of the scaling parameter $\alpha>1$. 

The first-order expansion of equation~\eqref{eq:solS0} yields
the following approximation of $\SZ$:

\begin{align}
  \SZ(t) & \simeq  1 - \epsilon \frac{t}{\taur} 
           +  \sum_{k=1}^{\floor{t/\taub}} \rho_{k}  \frac{t - k \taub}{\taur},
           \nonumber\\
         & =  \SZ\left(\floor{t/\taub}\taub\right) 
           - \frac{t-\floor{t/\taub}\taub}{\taur}
           \sum_{k = \ceiling{t/\taub}}^\infty \rho_{k} .
           \label{eq:solS0exp}
\end{align}

\noindent That is to say, for $\epsilon\ll1$, $\SZ(t)$ is a sequence of straight line
segments joining the values the function takes at integer multiples of
the propagation time $\taub$, 

\begin{align}
  \SZ(k \taub) 
  & 
    \simeq 1 - \epsilon \frac{\taub}{\taur} 
    \Big[ k - \sum_{j=1}^{k}   \frac{\rho_{j}}{\epsilon} (k-j) \Big]\,.
    \label{eq:S0expinteger}
\end{align}

\noindent In this regime, $\SZ(t)$ converges asymptotically to the constant
value \eqref{eq:S0largetime}, which, for the choice of parameters
\eqref{eq:rhok2}, is given by equation~\eqref{eq:S0rhok2}, i.e., with all
other parameters being fixed,

\begin{equation}
  \label{eq:S0largetimeepsi}
  \lim_{t\to\infty} \SZ(t) \simeq 1 -
  \epsilon (1 - 2^{1-\alpha})^{-1} \frac{\taub}{\taur}.  
\end{equation}

\noindent For these parameters, we make use of identities \eqref{eq:rhok2id}
to evaluate equation \eqref{eq:S0expinteger},

\begin{equation}
  \label{eq:S0expintegerepsi}
  \SZ(k \taub) \simeq 1 - \epsilon (1 - 2^{1-\alpha})^{-1}
  [1 - (k+1)^{1 - \alpha}] \frac{\taub}{\taur},
\end{equation}

\noindent or, for general time values,

\begin{align}
  \SZ(t) & \simeq 1 - \epsilon
           (1 - 2^{1-\alpha})^{-1}
           \Big\{ \frac{t}{\taur}  \Big[
           \big( \floor{t/\taub} + 1 \big)^{1 - \alpha}
           - \big( \floor{{t}/{\taub}} + 2\big)^{1 - \alpha}
           \Big]
           \cr
  & \quad
    + \frac{\taub}{\taur}  \Big[
    1 - \big( \floor{{t}/{\taub}} + 1\big)^{2 - \alpha}
    +  \big( \floor{{t}/{\taub}} + 2\big)^{2 - \alpha}
    \cr
  & \quad
    - 2\big( \floor{{t}/{\taub}} + 2\big)^{1 - \alpha}
    \Big]
    \Big\}.
    \label{eq:S0expepsi}
\end{align}

\noindent A comparison between this approximate value and the exact one,
equation \eqref{eq:solS0}, is displayed in \fref{fig:Szapprox}, where the
value of $\epsilon$ was taken to be large enough that
the curves remain distinguishable.

\section{Mean squared displacement}
\label{sec:msd}

Assuming initial injection of particles at the origin,
the mean squared displacement of particles as a function of time,
given by the second moment of the displacement vector $\mathbf{r}
\equiv \ell \mathbf{n}$, 
$\langle r^2 \rangle_t = \ell^2 \langle n^2
\rangle_t =  \ell^2 \sum_{\mathbf{n}\in\mathbb{Z}^d} n^2 P(\mathbf{n},t)$,
where $r^2 = \mathbf{r} \cdot \mathbf{r}$ and $n^2 = \mathbf{n} \cdot
\mathbf{n}$, evolves according to the  
differential equation 

\begin{align}
  \label{eq:dtn2a}  
  \td{t} \langle n^2 \rangle_t 
  & =  
    \taur^{-1} \sum_{k=0}^\infty (2k+1) \nu_k
    \SZ(t- k \taub)
    + 
    \sum_{k = 1}^\infty
    (2 k - 1)
    \sum_{k' = 0}^\infty 
    \sigma_{k + k'} (\mathbf{0}, t - k \taub)
    ,
\end{align}

\noindent 
where, in contrast to the expression derived in reference
\cite{Cristadoro:2014transport}, we have here added possible
contributions from source terms $\sigma_{k,j}(\mathbf{n}, t)$, $k\geq
1$, which we assume to be concentrated at site $\mathbf{n} =
\mathbf{0}$ only. To conform with stationarity of the process,
propagating states should be  uniformly injected in 
the time interval $-\taub < t \leq 0$ so they decay uniformly in the
time interval $0< t \leq \taub$. We thus let ($k \geq 1$)

\begin{equation}
  \label{eq:sigmak}
  \sigma_{k}(\mathbf{n}, t) 
  = \taub^{-1} \mu_k \delta_{\mathbf{n}, \mathbf{0}}  
  \Theta( - t)
  \Theta( t + \taub ),
\end{equation}

\noindent
where $\{\mu_k\}_{k\geq 1}$ is a sequence of positive numbers such
that $\sum_k \mu_k = 1$ and $\Theta(.)$ is the Heaviside step
function, such that the product $ \Theta( - t) \Theta( t + \taub )$
guarantees the uniform distribution of propagating source terms in the
desired time interval. Equation \eqref{eq:dtn2a} transforms to

\begin{equation}
  \label{eq:dtn2}
  \td{t} \langle n^2 \rangle_t 
  =  
  \taur^{-1} \sum_{k=0}^\infty (2k+1) \nu_k
  \SZ(t- k \taub)
  + \taub^{-1} 
  (2 \floor{t/\taub} + 1)\sum_{k = 1}^\infty
  \mu_{\floor{t/\taub} + k}
  .
\end{equation}

\noindent Integrating over time, we obtain the mean squared displacement:

\begin{align}
  \langle n^2 \rangle_t 
  & =  \frac{\taub}{\taur} \sum_{k = 0}^{\floor{t/\taub}} (2k+1) 
    \nu_k
    \int_0^{t/\taub - k }\dd{x} \SZ(x \taub)
    + 
    \sum_{k = 1}^\infty
    \int_0^{t/\taub}
    \dd{x} 
    (2 \floor{x} + 1)
    \mu_{\floor{x} + k}
    .
    \label{eq:n2}
\end{align}

\noindent
There are two contributions to this expression; the first
arises from the fraction of particles in a scattering state,
and the second from particles initially distributed among
propagating states. This term transforms to

\begin{align}
  \sum_{k = 1}^\infty
  \int_0^{t/\taub}
  \dd{x} 
  (2 \floor{x} + 1)
  & \mu_{\floor{x} + k}
    = 
    \sum_{k = 0}^\infty
    \sum_{j = 1}^{\floor{t/\taub}} (2 j - 1) \mu_{j + k}
    \cr
  &
    + 
    2 ( t/\taub - \floor{t/\taub} )
    (\floor{t/\taub} + 1/2) \sum_{k = 1}^\infty \mu_{\floor{t/\taub} +
    k}
    \,.
    \label{eq:n2prop}
\end{align}

\subsection{Stationary regimes \label{sec:n2asym}}

The \emph{stationary initial condition}, i.e.~such that the initial fractions
of particles in scattering and propagating states are stationary
solutions of equations \eqref{eq:dtP0}-\eqref{eq:dtPkj}, are realized if,
in equation \eqref{eq:sigmak}, we identify $\mu_k$ with the stationary
fraction of particles in state $k$, 
\begin{equation}
  \label{eq:mukss}
  \mu_k = \frac{\nu_k \taub}
  {\taur + \epsilon (1 - 2^{1-\alpha})^{-1} \taub},
\end{equation}
cf.~equation \eqref{eq:Skrhok2}, and let
\begin{equation}
  \label{eq:mu0ss}
  \sigma_0(\mathbf{n}, t) = \mu_0
  \deltad(t) \delta_{\mathbf{n}, \mathbf{0}}, 
  \qquad
  \mu_0 = 
  \frac{\taur}
  {\taur + \epsilon (1 - 2^{1-\alpha})^{-1} \taub}.
\end{equation}

\noindent With this choice, the mean squared displacement \eqref{eq:n2}
yields the exact expression

\begin{align}
  \langle n^2 \rangle_t 
  & =     
    \frac{t} {(1 - 2^{1-\alpha}) \taur + \epsilon \taub} 
    \bigg\{
    1 - 2^{1-\alpha} + 
    \frac{\epsilon}{t}  \sum_{k = 1}^{\floor{t/\taub}}
    (t - k \taub) (2 k + 1) 
    [k^{1-\alpha} - 
    (k+1)^{1-\alpha}]
    \bigg\}
    \cr
  &
    \quad
    + \frac{\epsilon \taub} 
    {(1 - 2^{1-\alpha}) \taur + \epsilon \taub}     
    \bigg\{
    \sum_{k=1}^{\floor{t/\taub}} 
    \frac{2k - 1} {k^{\alpha - 1}}
    + 2 ( t/\taub - \floor{t/\taub} )
    \frac{\floor{t/\taub} + 1/2}
    { ( \floor{t/\taub} + 1)^{\alpha - 1}}
    \bigg\}
    \,.
    \label{eq:n2ss}
\end{align}

\noindent
Treating $n \equiv t/\taub \gg 1$ as an integer and 
letting $H_n^{(\beta)} \equiv \sum_{k=1}^n k^{-\beta}$ denote the  generalized
harmonic numbers, the
two terms with sums over $k$ in the above expression evaluate
respectively to   

\begin{align}
  \sum_{k = 1}^n
  n^{-1} (n - k ) (2 k + 1) 
  [k^{1-\alpha} - 
  (k+1)^{1-\alpha}]
  \simeq 1 - 4 n^{-1} H_n^{(\alpha -2)}
  + 2 H_n^{(\alpha -1)}\,,
  \label{eq:n2asym_evalss}
\end{align}

\noindent for the term arising from the fraction of particles in a
scattering state, and to

\begin{equation}
  \label{eq:n2asym_evalps}
  \sum_{k=1}^{n} (2k - 1) k^{1 - \alpha} = 2 H_n^{(\alpha -2)} -
  H_n^{(\alpha -1)}\,,
\end{equation}

\noindent  for the term arising from particles initially
distributed among the propagating states. Depending on the value of
the scaling parameter $\alpha$, we have the following three asymptotic 
regimes.

\subsubsection{\textsc{Normal diffusion}  \label{sec:n2asym_agt2}}

For parameter values  $\alpha > 2$, the asymptotic properties of the harmonic numbers,

\begin{equation}
  \label{eq:Hn_agt2}
  \begin{split}
    &\lim_{n\to\infty} n^{-1} H_n^{(\alpha - 2)} = 0,\\  
    &\lim_{n\to\infty} H_n^{(\alpha - 1)} = \zeta(\alpha - 1),
  \end{split}
\end{equation}

\noindent are such that 
only the terms arising from the fraction of particles in a
scattering asymptotically contribute to equation \eqref{eq:n2ss}:

\begin{equation}
  \lim_{t\to\infty} \frac{1}{t} \langle n^2 \rangle_t
  =
  \frac{1 - 2^{1-\alpha} + \epsilon 
    [ 1  + 2 \zeta(\alpha - 1)]}
  {(1 - 2^{1-\alpha})\taur + \epsilon \taub} ,
  \label{eq:msdasympt_agt2}
\end{equation}

\noindent which, up to a factor $z^{-1}$, is the diffusion coefficient of the
process.

\subsubsection{\textsc{Weak superdiffusion} \label{sec:n2asym_aeq2}}

For the marginal parameter value $\alpha = 2$, the harmonic numbers in
equation \eqref{eq:n2asym_evalss} evaluate to 

\begin{equation}
  \label{eq:Hn_aeq2}
  \begin{split}
    &H_n^{(0)} = n,\\  
    &H_n^{(1)} = \log n + \gamma + \Or(n^{-1}),
  \end{split}
\end{equation}

\noindent where $\gamma \simeq 0.577216$ is Euler's constant.

We therefore have the asymptotic limit of equation \eqref{eq:n2ss},

\begin{equation}
  \lim_{t\to\infty} \frac{1}{t \log (t/\taub)} \langle n^2 \rangle_t 
  =
  \frac{4 \epsilon}
  {\taur + 2\epsilon \taub} \,,
  \label{eq:msdasympt_aeq2}
\end{equation}

\noindent 
which is due to the fraction of particles in a scattering state alone.
This result is, however, of limited use because the asymptotic regime
 only emerges provided $\epsilon \log (t/\taub) \gg1$
(where $1$ corresponds to the order of the sub-leading term), which may
not be attainable, especially when the scattering parameter is small.
For this reason, it is preferable to retain also the next-order term in
equation \eqref{eq:n2ss},

\begin{align}
  \frac{1}{t} \langle n^2 \rangle_t 
  & \simeq
    \frac{1}
    {\taur + 2 \epsilon \taub} 
    \Big\{ 1
    + 4 \epsilon  \big[ \log (t/\taub) +
    \gamma 
    - 1/2
    + \Or(t/\taub)^{-1} \big] 
    \Big\}\,,
    \label{eq:msdasymptsim_aeq2}
\end{align}

\noindent 
where a fraction $4 \epsilon/(\taur + 2 \epsilon \taub)$ is
contributed by particles initially distributed among propagating
states.

\subsubsection{\textsc{Superdiffusion}  \label{sec:n2asym_alt2}} 

In the range of parameters $1<\alpha<2$, we substitute the scaling
properties of the harmonic numbers,  

\begin{equation}
  \label{eq:Hn_alt2}
  \begin{split}
    &\lim_{n\to\infty} n^{ \alpha - 3} H_n^{(\alpha - 2)} = (3 - \alpha)^{-1},\\  
    &\lim_{n\to\infty} n^{\alpha - 2} H_n^{(\alpha - 1)} = (2 - \alpha)^{-1}.
  \end{split}
\end{equation}

\noindent and obtain from equation \eqref{eq:n2ss} the limit

\begin{align}
  \lim_{t\to\infty} \frac{1}{(t/\taub)^{3 - \alpha}} 
  \langle n^2 \rangle_t
  & =
    \left[
    \frac{\alpha - 1}{(2 - \alpha)(3 - \alpha)} 
    + 
    \frac{1}{3 - \alpha}
    \right]
    \frac{ 2\epsilon \taub}
    {(1 - 2^{1-\alpha})\taur + \epsilon \taub} \,,
    \cr
  & =
    \frac{1}{(2 - \alpha)(3 - \alpha)} 
    \frac{ 2\epsilon \taub}
    {(1 - 2^{1-\alpha})\taur + \epsilon \taub} \,.
    \label{eq:msdasympt_alt2}
\end{align}

\noindent 
As emphasized by the first line of this equation, the leading-order
coefficient is the sum of two distinct non-trivial contributions from
the two terms on the right-hand side of equation \eqref{eq:n2ss}: one
from the fraction of particles in a scattering state and the other
from particles initially injected in propagating states. As a
consequence, the transport coefficient corresponding to the
all-scattering initial condition differs by a factor $\alpha - 1$ from 
that above, corresponding to the stationary initial condition. This
observation is consistent with the results of reference 
\cite{Zumofen:1993PowerSpectra}, where non-recurrent regimes of
the scaling parameter $\alpha$ were also investigated.

\subsection{Exact time-dependent solution \label{sec:n2sol}}

The following exact expression of the mean squared displacement
\eqref{eq:n2} is obtained after substitution of the fraction of
particles in a scattering state, equation \eqref{eq:solS0}, 
and  applies to the all-scattering initial condition \eqref{eq:sigma0}:

\begingroup
\allowdisplaybreaks
\begin{align}
  \langle n^2 \rangle_t 
  &=  \sum_{k = 0}^{\floor{t/\taub}} (2k+1) 
    \sum_{l = k}^{\infty} \epsilon^{-1}\rho_{l}  
    \Bigg\{
    1 - e^{-\epsilon (t-k \taub)/\taur} 
    \nonumber\\
  & \quad
    + \sum_{j=1}^{\floor{t/\taub}-k-1} 
    \sum_{m=1}^j 
    \sum_{n=1}^m 
    \epsilon^{-n} a_{(n|m)}  \,
    e^{-\epsilon(j-m)\taub/\taur} \sum_{i=0}^{n} \frac{n!}{i!}  
    \epsilon^{i} \left(\frac{\taub}{ \taur}\right)^{i} 
    \nonumber\\
  & \qquad
    \times 
    \Big[(j-m)^{i} - (j-m+1)^{i} e^{-\epsilon\taub/\taur}\Big]
    \nonumber\\
  & \quad
    + \sum_{m=1}^{\floor{t/\taub}-k} \sum_{n=1}^m 
    \epsilon^{-n} a_{(n|m)} \,
    e^{-\epsilon (\floor{t/\taub}-k-m) \taub/\taur}
    \sum_{i=0}^{n} \frac{n!}{i!} \epsilon^{i}
    \left(\frac{\taub}{\taur}\right)^{i} 
    \nonumber\\
  & \qquad 
    \times 
    \Big[(\floor{t/\taub}-k-m)^{i} - (t/\taub-k-m)^{i}
    e^{-\epsilon(t-\floor{t/\taub} \taub)/\taur}\Big]
    \Bigg\},
    \label{eq:n2sol}
\end{align}
\endgroup
\noindent where, for consistency, one must interpret terms such as
$(j-m)^0$ as unity, even when $j = m$. 

\begin{figure}[h]
  \centering
  \begin{subfigure}[b]{0.49\textwidth}
    \includegraphics[width=\textwidth]
    {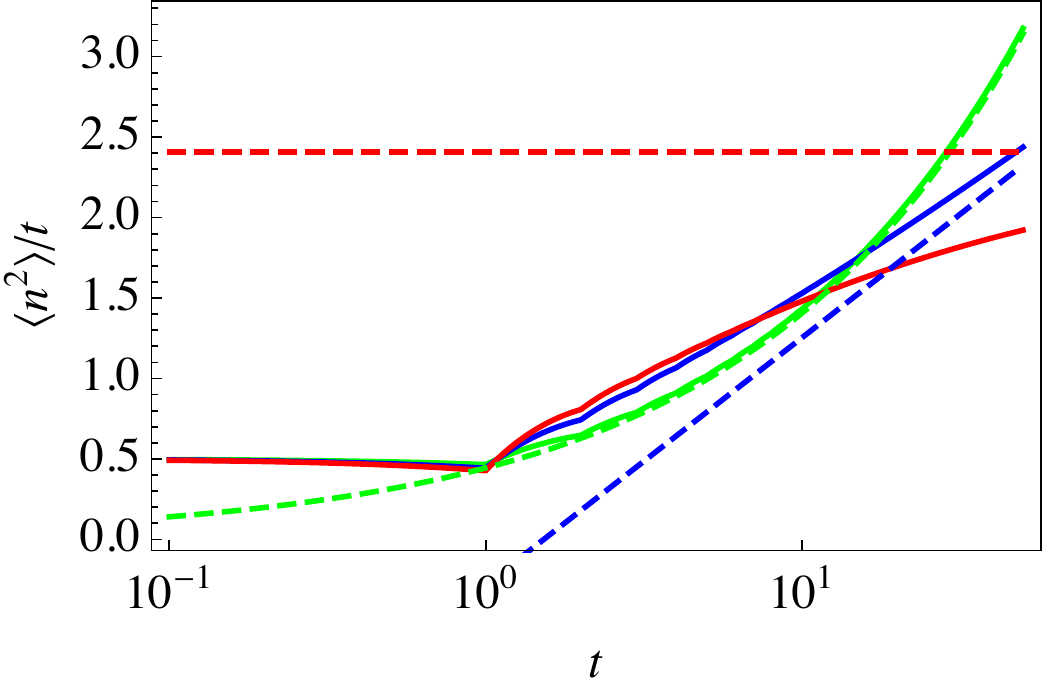}
    \caption{
      $\epsilon(1 - 2^{1-\alpha})^{-1} = 1$} 
    \label{fig:msdexact}
  \end{subfigure}
  \begin{subfigure}[b]{0.49\textwidth}
    \includegraphics[width=\textwidth]
    {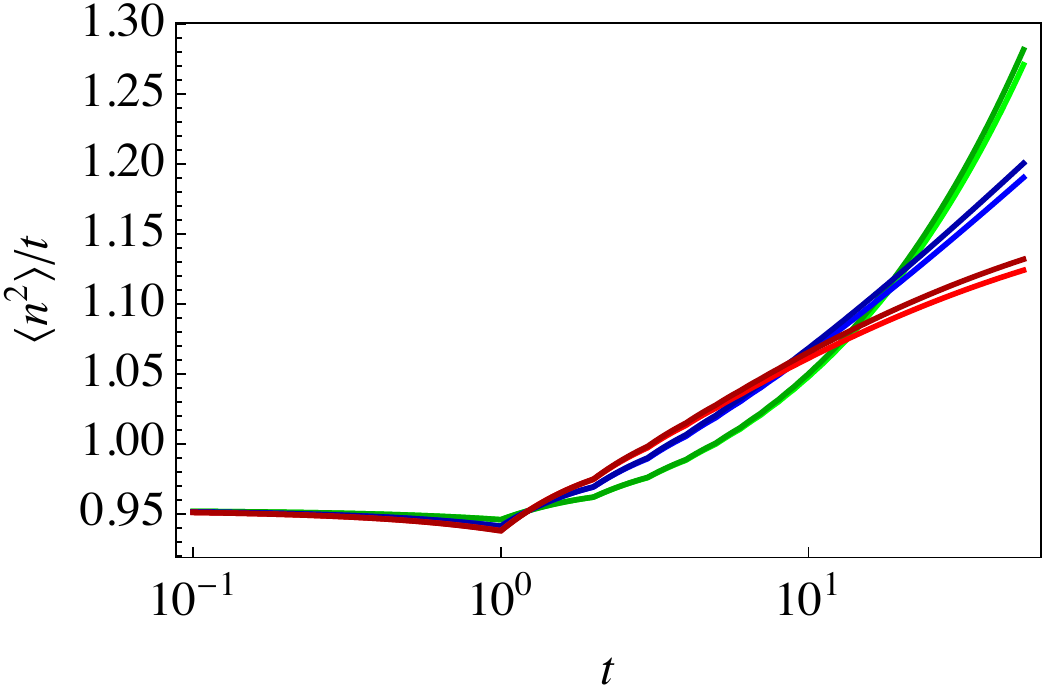}
    \caption{
      $\epsilon(1 - 2^{1-\alpha})^{-1} = 5\times 10^{-2}$} 
    \label{fig:msdepsi}
  \end{subfigure}
  \caption{
    Growth in time of the rescaled mean squared
    displacement, $\langle n^2 \rangle_t/t$, computed from
    equation \eqref{eq:n2sol}. The transition probabilities were set
    according to equations \eqref{eq:rhok2}, with $\taub = 1$
    and $\taur = 1 + \epsilon(1 - 2^{1 - \alpha})^{-1}$.
    The value of the scattering parameter was
    chosen such that $\epsilon/(1 - 2^{1-\alpha}) = 1$ (left panel)
    and $5 \times 10^{-2}$ (right panel). In both figures, the scaling
    parameter takes the values $\alpha = 3/2$ (green 
    curve), $2$ (cyan curve), and $5/2$ (red curve).      
    For comparison, in the left panel, the asymptotic scaling
    values predicted by equations \eqref{eq:msdasympt_agt2},
    \eqref{eq:msdasympt_aeq2} and 
    \eqref{eq:msdasympt_alt2} are displayed in dashed lines with
    matching colors. The right panel shows a comparison between 
    the exact solution and an approximated solution, exact to first
    order in $\epsilon$ (darker curves); see equation
    \eqref{eq:n2epsitime}.
  }
  \label{fig:msd}
\end{figure}

Figure \ref{fig:msd} displays the results of explicit evaluations
of equation \eqref{eq:n2sol} for three different values of the scaling
parameter, $\alpha>1$, with $\epsilon = 1 - 2^{1-\alpha}$. According to
the asymptotic results \eqref{eq:msdasympt_agt2},
\eqref{eq:msdasympt_aeq2} and \eqref{eq:msdasympt_alt2}, the scaling
parameter value $\alpha =2$ gives rise to a logarithmic divergence of
$\langle n^2 \rangle_t/t$, which separates the superdiffusive regime,
for $1<  \alpha <2$, from that of normal diffusion, for
$\alpha>2$.  A comparison between the exact solution and the
asymptotic ones is displayed in \fref{fig:msdexact}. On the time scale
of the figure, convergence to the 
asymptotic regimes is convincingly observed only in the superdiffusive
case.  Note that while the computation of equation \eqref{eq:n2sol} up
to $t = 50 \taub$ takes less than one half hour of CPU time on a
reasonably fast computer, doubling the time range would increase the
required CPU time to over ten days. 

Although equation \eqref{eq:n2sol} gives the exact time-dependence of
the mean squared displacement of the process for the initial condition
\eqref{eq:S0ic}, it does not give much insight into its time
development. In particular, how  to extract the large-time behaviour
of the mean squared displacement and connect this result to the
asymptotic scalings \eqref{eq:msdasympt_agt2}, 
\eqref{eq:msdasympt_aeq2} and \eqref{eq:msdasympt_alt2} is not
transparent. As reflected by \fref{fig:msdexact}, a numerical
evaluation of equation \eqref{eq:n2sol} is necessarily limited to
moderately large times, due to the number of terms involved.

A regime of specific interest which allows us to infer the emergence
of the asymptotic scalings \eqref{eq:msdasympt_agt2}, 
\eqref{eq:msdasympt_aeq2} and \eqref{eq:msdasympt_alt2} from the
solution \eqref{eq:n2sol} is, however, that of small values of the
scattering parameter $\epsilon$, i.e., such that transitions from
scattering to propagating states occur only rarely. This is discussed
below.  We will otherwise have to resort to
numerical simulations of the underlying stochastic processes to
observe this convergence. Those results are presented in 
\sref{sec:numn2}.

\subsection{Small-parameter expansion \label{sec:n2exp}}

For small values of the scattering parameter $\epsilon$, when
transitions from scattering to propagating states are much rarer than
transitions between scattering states, an expansion of equation
\eqref{eq:n2sol} in powers of this parameter provides an approximate
expression of the mean squared displacement from which the different
asymptotic regimes discussed in \sref{sec:n2asym} can be inferred. 
In the anomalous regime of the scaling parameter, $\alpha \geq 2$,
terms constant in time, that carry a normal contribution to the mean
squared displacement, may bring 
about contributions which, even for large times, may be much larger
than that of terms diverging in time; an anomalous
contribution to the mean squared displacement may then be masked by a
normal one.  

To proceed, we substitute equation \eqref{eq:S0expintegerepsi} into
equation \eqref{eq:n2} and let $t \equiv k \taub$. Having dropped all terms
of order $\epsilon^2$ and higher, we obtain

\begin{align}
  \langle n^2 \rangle_{k \taub} 
  & =  
    \frac{\taub}{\taur} 
    \int_0^{k}\dd{x} 
    \Big\{1 - \epsilon (1 - 2^{1-\alpha})^{-1} \frac{\taub}{\taur}
    [1 - ( \floor{x}+1)^{1 - \alpha}] 
    \cr
  &\quad
    - \frac{\taub}{\taur}
    (x - \floor{x})
    \sum_{i = \ceiling{x}}^\infty \rho_i
    \Big\}
    +
    \frac{\taub}{\taur} \sum_{j = 1}^{k} (2j+1) 
    (k - j) \sum_{i = j}^{\infty}\rho_{i} 
    \cr
  & = 
    \frac{k \taub}{\taur} 
    \Big\{
    1 - 
    \frac{\taub}{\taur} \epsilon (1 - 2^{1-\alpha})^{-1}
    \big[ 1 - k ^{-1} (H_k^{(\alpha - 1)} - 1/2 + 1/2(k+1)^{1 - \alpha}  )
    \big]
    \cr
  &\quad
    + \epsilon (1 - 2^{1-\alpha})^{-1}
    \Big[
    1 - (k + 1)^{1 - \alpha}
    + (2 - k^{-1})( H_{k+1}^{(\alpha - 1)} - (k + 1)^{2 - \alpha} )
    \cr
  &\quad
    - 2 k^{-1} ( 2 H_{k+1}^{(\alpha - 2)} - H_{k+1}^{(\alpha - 1)} -
    (k+1)^{3 - \alpha} )
    \Big]
    \Big\}\,,
    \cr
  &\simeq
    \frac{k \taub}{\taur} 
    \Big\{
    1 + \epsilon (1 - 2^{1-\alpha})^{-1}
    \Big[ 1 + 
    2 H_{k}^{(\alpha - 1)} - 4 k^{-1}H_{k}^{(\alpha - 2)} 
    - 
    \frac{\taub}{\taur}
    \Big]
    \Big\}\,,
    \label{eq:n2epsi}
\end{align}

\noindent 
where, in the last line, we have omitted terms that decay to zero as 
$k$ grows large. Plugging into this expression the asymptotic
forms of the generalized harmonic numbers \eqref{eq:Hn_agt2},  
\eqref{eq:Hn_aeq2} and \eqref{eq:Hn_alt2}, 

\begin{equation}
  \label{eq:Hn_a}
  H_{k}^{(\alpha - 1)} - 2 k^{-1}H_{k}^{(\alpha - 2)} 
  \simeq 
  \begin{cases}
    \zeta(\alpha - 1),
    &\alpha>2,\\ 
    \log k + \gamma - 2,
    &\alpha=2,\\ 
    \frac{\alpha - 1}{(2 - \alpha)(3 - \alpha)}k^{2 - \alpha} + \zeta(\alpha - 1),
    &1<\alpha<2,
  \end{cases}
\end{equation}

\noindent we obtain approximations of the mean squared displacement
\eqref{eq:n2sol}, which can be compared with the asymptotic expressions
\eqref{eq:msdasympt_agt2}, \eqref{eq:msdasympt_aeq2} and
\eqref{eq:msdasympt_alt2}.

In the regime of normal diffusion, $\alpha >2$, equation
\eqref{eq:n2epsi} reduces to the first order expansion in $\epsilon$ of
equation \eqref{eq:msdasympt_agt2}. For $1< \alpha \leq 2$, the mean
squared displacement \eqref{eq:n2epsi} displays normal 
diffusion at short times and anomalous diffusion at large times. Thus
in the weak superdiffusive regime, $\alpha = 2$, equation
\eqref{eq:n2epsi} reduces to 

\begin{equation}
  \frac{1}{t} \langle n^2 \rangle_t 
  \simeq  
  \frac{1}{\taur}
  \Big\{ 1+ 
  2 \epsilon 
  \Big[ 2 \log \frac{t}{\taub} + 2 \gamma - 3 - \frac{\taub}{\taur}
  \Big]
  \Big\},
  \label{eq:n2epsi_aeq2}
\end{equation}

\noindent 
which differs from the stationary expression
\eqref{eq:msdasymptsim_aeq2} by  a factor $4\epsilon/\taur$, 
subleading with respect to the logarithmically diverging term,
identical in both expression. This difference stems from the absence
of particles populating propagating states in our choice of initial
condition.
In the superdiffusive regime, $1 < \alpha < 2$, the divergence of the
harmonic numbers, equation \eqref{eq:Hn_a}, dominates the large time
behaviour,

\begin{equation}
  \frac{1}{t} \langle n^2 \rangle_t 
  \simeq
  \frac{1}{\taur}
  \Big\{ 1+ 
  \epsilon (1 - 2^{1-\alpha})^{-1}
  \Big[ \frac{2(\alpha - 1)}{(2 - \alpha)(3 - \alpha)} \Big(
  \frac{t}{\taub}\Big)^{2 - \alpha} 
  + 1 + 2 \zeta(\alpha - 1) - \frac{\taub}{\taur}
  \Big]
  \Big\}.
  \label{eq:n2epsi_alt2}
\end{equation}

\noindent
Here again we note a difference between this expression and equation
\eqref{eq:msdasympt_alt2}, obtained in the stationary regime. At
variance with the case $\alpha = 2$, however, the difference occurs in the
coefficient of the leading term on the right-hand side of equation
\eqref{eq:n2epsi_alt2}, whose numerator is $1- \alpha$ instead of
$1$ in the stationary regime. The transport coefficient on the
right-hand side of equation \eqref{eq:n2epsi_alt2} therefore
changes, depending on the choice of initial conditions.

At large times, after the mean squared displacement crosses over from
normal diffusion to superdiffusion, the above expressions are
equivalent to $\Or(\epsilon)$ expansions of the asymptotic
values \eqref{eq:msdasympt_aeq2} and \eqref{eq:msdasympt_alt2}. The
value of the crossover time, $t_\mathrm{c}$, that separates the short
time normal diffusion from the large time anomalous diffusion can be
inferred from the above  expressions:

\begin{equation}
  \label{eq:timesupdiff}
  t_\mathrm{c}
  \approx \taub
  \times
  \begin{cases}
    \left[
      \frac{(2 - \alpha)(3 - \alpha)}
      {2(\alpha - 1)} 
      \frac{1 - 2^{1 - \alpha}}
      {\epsilon}
    \right]^{\frac{1}{2 - \alpha}}, &
    1 < \alpha < 2,\\
    \exp \left( 
      \frac{1}{4\epsilon}
    \right),
    &
    \alpha =2,
  \end{cases}
\end{equation}

\noindent which can be large, in particular when $\alpha = 2$.

For short times, equation \eqref{eq:n2epsi} can be improved by
removing the assumption that $t$ is an integer multiple of
$\taub$. One then finds

\begin{align}
  \langle n^2 \rangle_t &=
                          \frac{t}{\taur} + \epsilon (1 - 2^{1-\alpha})^{-1}
                          \bigg\{
                          \cr
  &\quad
    \frac{t}{\taur} 
    \Big[
    1 - 2 (\floor{t/\taub} + 1)^{2 - \alpha} 
    - (\floor{t/\taub} + 1)^{1 - \alpha}
    + 2 H_{\floor{t/\taub} + 1}^{(\alpha - 1)}
    \Big]         
    \cr
  & \quad
    + \frac{\taub}{\taur}
    \Big[
    2 (\floor{t/\taub} + 1)^{3 - \alpha}
    + (\floor{t/\taub} + 1)^{2 - \alpha}
    - 4 H_{\floor{t/\taub} + 1}^{(\alpha - 2)}
    + H_{\floor{t/\taub} + 1}^{(\alpha - 1)}
    \Big]
    \cr
  & \quad
    - \left( \frac{\taub}{\taur} \right)^2
    \Big\{
    \floor{t/\taub} - H_{\floor{t/\taub}}^{(\alpha - 1)} + 1/2 - 1/2
    (\floor{t/\taub} + 1)^{1 - \alpha}
    \cr
  & \qquad
    + 1/2 \big[ (t/\taub)^2 - \floor{t/\taub}^2 \big]
    \big[ (\floor{t/\taub} + 1)^{1 - \alpha} 
    - (\floor{t/\taub} + 2)^{2 - \alpha} \big]
    \cr
  & \qquad
    + \big( t/\taub - \floor{t/\taub} \big)
    \big[
    1 - (\floor{t/\taub} + 1)^{2 - \alpha} 
    + (\floor{t/\taub} + 2)^{2 - \alpha}
    \cr
  & \qquad
    -2 (\floor{t/\taub} + 2)^{1 - \alpha}
    \big]
    \Big\}
    \bigg\}\,.
    \label{eq:n2epsitime}
\end{align}

\noindent 
A comparison between this approximate solution and the exact one,
equation \eqref{eq:n2sol}, is shown in \fref{fig:msdepsi} for
different values of the scaling parameter, $\alpha$.

\section{Numerical computations}
\label{sec:numn2}

Numerical simulations of the process with rates \eqref{eq:psik} and
probabilities \eqref{eq:Pkj} are based on a classic kinetic Monte
Carlo algorithm \cite{Gillespie:1976p296}, taking into account the
possibility of ballistic motion of particles in a propagating phase.

A collection of independent walkers are initialized at time $t=0$ at
the origin of the two-dimensional square lattice $\mathbb{Z}^2$,
either in the scattering state, $k_0=0$, for the all-scattering initial
condition, or in state $k_0 \geq 0$ with relative weights $\mu_{k_0}$
specified by equations  \eqref{eq:mukss}--\eqref{eq:mu0ss}, for
the stationary initial condition. For each walker, we generate a sequence 
$\{(k_n, j_n), t_n\}_{n\in\mathbb{N}}$ of successive states $(k_n,
j_n)$, $k_n \in \mathbb{N}$, $j_n = 1, \dots, z \equiv 4$, and
corresponding times $t_n$ as follows.

When in a state $k_{n-1} = 0$, the next transition is determined by drawing
the following three random numbers. This is referred to as a
\emph{scattering step}. 
\begin{itemize}
\item[(i)]
  The first random number, $\eta \in [0,1]$, is
  drawn from a uniform 
  distribution, and yields the state $k_{n}$, such that
  $\sum_{a = 0}^{k_{n}-1} \rho_a \leq \eta <   
  \sum_{a = 0}^{k_{n}} \rho_a$. 
\item[(ii)]
  The corresponding waiting time, i.e., the time $t_{n}$ the
  particle waits in the state $k_{n-1} = 0$ before the transition to state
  $k_{n}$ takes place, is obtained by drawing a second random
  number, whose distribution is exponential with mean $\taur$. 
\item[(iii)]
  A third random integer, with uniform distribution among the set of
  lattice directions $\{1,\dots,4\}$, identifies the direction $j_{n}$
  of the corresponding displacement by one lattice unit. 
\end{itemize}

Depending on the new state $k_{n}$, another scattering step is taken
if $k_{n} = 0$ or, if $k_{n}\geq 1$, a sequence of $k_{n}$ \emph{propagating
  steps} $\{(k_m, j_m), t_m\}_{m = n+1}^{n + k_n}$ takes place, such
that: 
\begin{itemize}
\item[(i)]
  the state decreases by one unit, $k_m = k_{m-1} - 1$,
\item[(ii)]
  the direction remains unchanged
  $j_m =j_{m-1}$, and 
\item[(iii)] 
  the time step takes value $t_m = \taub$, corresponding
  to the propagation time over a single lattice cell.
\end{itemize}
The above loop repeats itself until $m = n + k_n$, that is until
$k_{m} = 0$, at which point another scattering step is taken.

Keeping track of the positions of all walkers as functions of time,
which we measure at regular intervals on a logarithmic timescale,
measurements of the mean squared displacement $\langle n^2 \rangle_t$
rescaled by $1/t$ are performed by taking averages of this quantity
over the set of all walkers, which typically consists of $10^8$
trajectories. Such measurements are reported  below for the different
scaling regimes in the positive recurrent range of the scaling
parameter values, $\alpha > 1$.

Throughout this section, we set the transition probabilities according to
equation \eqref{eq:rhok2} and, for convenience, change the scattering
parameter $\epsilon$ to  

\begin{equation}
  \label{eq:numdelta}
  \delta \equiv \frac{2 \epsilon} {1 - 2^{1-\alpha} + \epsilon}\,.
\end{equation}

\noindent We further let

\begin{equation}
  \label{eq:numtaubtaur}
  \begin{split}
    \taub &\equiv 1\,,\\
    \taur &\equiv \frac{2}{2 - \delta}\,,
  \end{split}
\end{equation}

\noindent such that the asymptotic fraction of
particles in the scattering state \eqref{eq:S0rhok2} becomes

\begin{equation}
  \lim_{t\to\infty} \SZ(t) = \frac{2}
  {2 + \delta}
  \, .
  \label{eq:numS0}
\end{equation}

\subsection{All-scattering vs.~stationary initial
  conditions  \label{sec:numssvas}}

With the choice of parameters
\eqref{eq:numdelta}-\eqref{eq:numtaubtaur}, The mean squared
displacement, for $t$ large and with the all-scattering initial
condition, is expected to scale as 

\begin{equation}
  \label{eq:numn2}
  \frac{\langle n^2 \rangle_t}{t}
  \simeq
  \frac{2}{2 + \delta}
  \times
  \begin{cases}
    1 + \delta \, \zeta(\alpha -1) \,,
    & \alpha >2\,,\\
    1 + \delta ( \log t + \gamma - 2)\,,
    & \alpha = 2\,,\\
    1 + \delta \Big[
    \frac{\alpha - 1}{(2 - \alpha) (3 - \alpha) }
    \, t^{2 - \alpha} 
    + \zeta(\alpha - 1)\Big]\,,
    & 1 < \alpha < 2\,,
  \end{cases}
  \quad \mbox{(all-scattering i.~c.)}
\end{equation}

\noindent 
where, in the two anomalous cases, we kept terms constant in time to
reflect the possibility that, when $\delta$ is small, the normal
term may not be negligible with respect to the anomalous one
over a large range of times. This is particularly relevant for the
marginal case, $\alpha  = 2$, where $\log t$ and $\gamma - 2$ remain
of comparable sizes throughout the time range accessible to numerical
computations. In comparison, for the stationary initial condition, the
above expression remains unchanged in the 
normal diffusive case, but is modified in the two anomalous cases, 

\begin{equation}
  \label{eq:numn2ss}
  \frac{\langle n^2 \rangle_t}{t}
  \simeq
  \frac{2}{2 + \delta}
  \times
  \begin{cases}
    1 + \delta \, \zeta(\alpha -1) \,,
    & \alpha >2\,,\\
    1 + \delta ( \log t + \gamma - 1)\,,
    & \alpha = 2\,,\\
    1 + \delta \Big[
    \frac{1}{(2 - \alpha) (3 - \alpha) }
    \, t^{2 - \alpha} 
    + \zeta(\alpha - 1)\Big]\,,
    & 1 < \alpha < 2\,.
  \end{cases}
  \quad \mbox{(stationary i.~c.)}
\end{equation}

\noindent The first order expansion in $\delta$ of the two
anomalous regimes in equation \eqref{eq:numn2} (in all-scattering
initial condition) yields results equivalent to equations
\eqref{eq:n2epsi_aeq2} and 
\eqref{eq:n2epsi_alt2} respectively.

\begin{figure}[p]
  \centering
  \begin{subfigure}[b]{0.5\textwidth}
    \includegraphics[width=\textwidth]
    {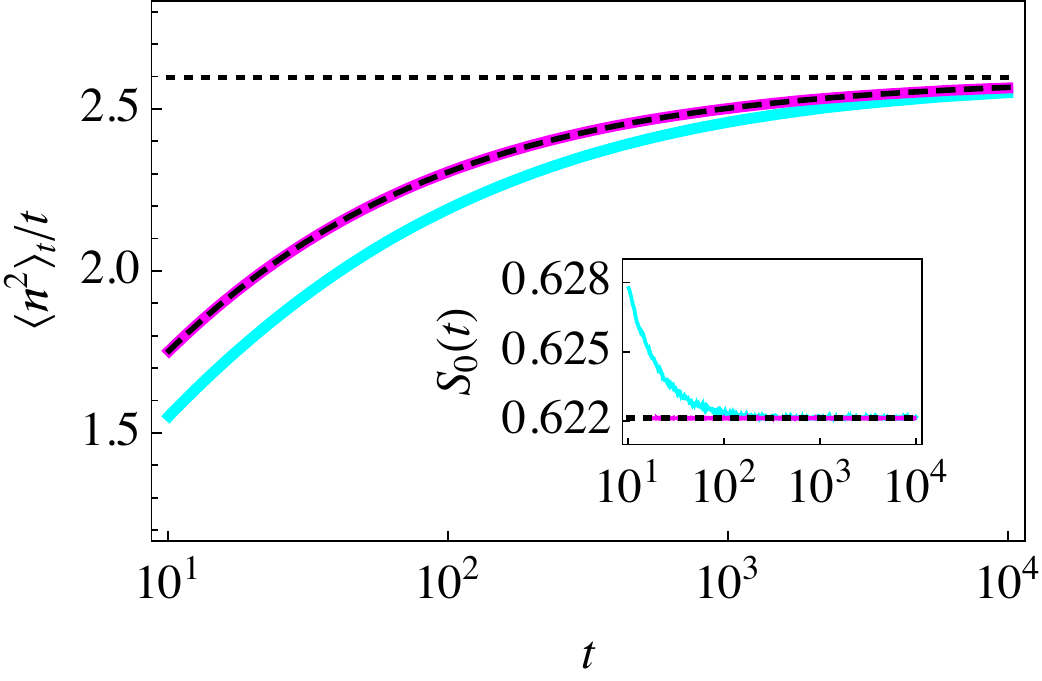} 
    \caption{$\alpha = 5/2$, $\delta \simeq 1.215$ ($\epsilon = 1$) } 
    \label{fig:msdss_alphagt2}
  \end{subfigure}
  \begin{subfigure}[b]{0.5\textwidth}
    \includegraphics[width=\textwidth]
  {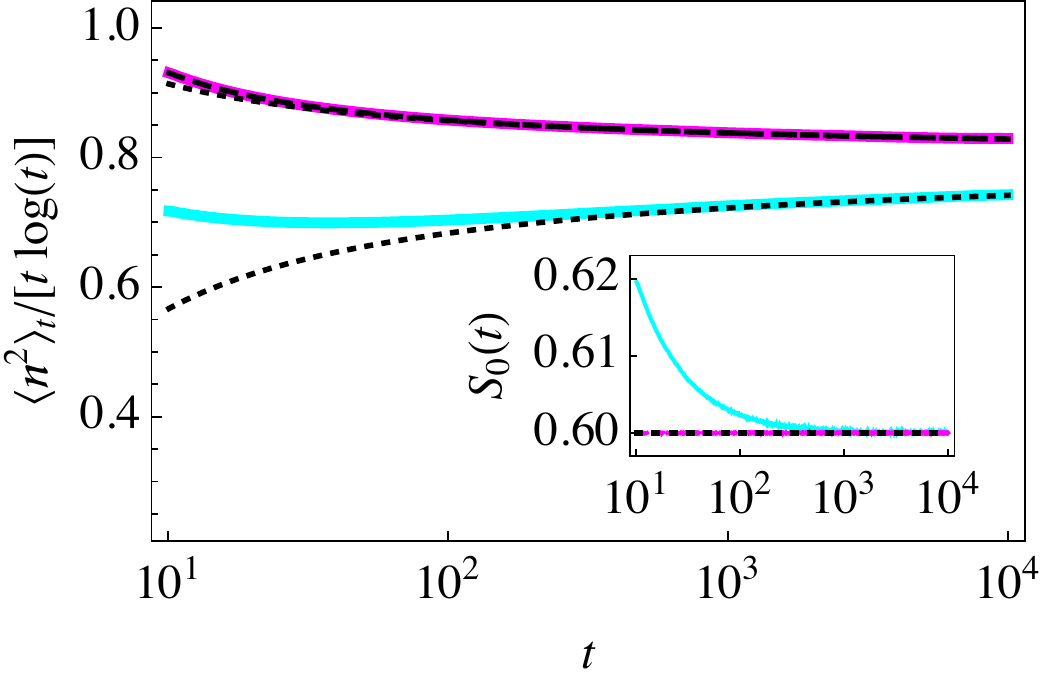} 
    \caption{$\alpha = 2$, $\delta = 4/3$ ($\epsilon = 1$) }
    \label{fig:msdss_alphaeq2}
  \end{subfigure}
  \begin{subfigure}[b]{0.5\textwidth}
    \includegraphics[width=\textwidth]
    {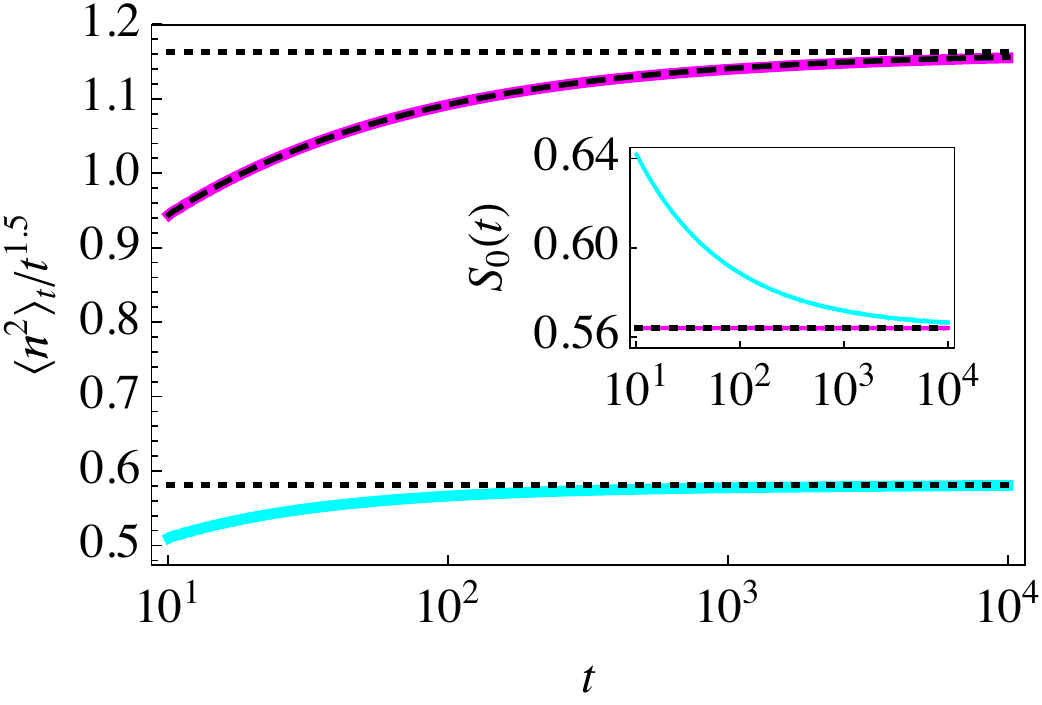} 
    \caption{$\alpha = 3/2$, $\delta \simeq 1.547$ ($\epsilon = 1$) }
    \label{fig:msdss_alphalt2}
  \end{subfigure}
  \caption{Numerical measurement of the time-evolution of the normally
    or anomalously rescaled mean squared displacement in the three
    distinct regimes of the scaling parameter. Time evolutions obtained
    from the stationary initial condition (magenta curves) are compared
    with those generated by the all-scattering initial condition (cyan
    curves). The dashed black curves correspond to the exact result
    \eqref{eq:n2ss}  and the dotted lines to the asymptotic regimes
    \eqref{eq:numn2} and \eqref{eq:numn2ss}, identical for regimes of
    normal diffusion, but otherwise differing according to 
    the type of initial conditions. Insets: time evolution of the
    fraction of particles in the scattering state compared with the
    stationary value \eqref{eq:numS0} (black dotted line). 
  }
  \label{fig:msdss_alpha}
\end{figure}

A computation of the time evolution of the mean squared displacement
in regimes of normal ($\alpha > 2$) and anomalous ($1 < \alpha \leq
2$) diffusion is shown in \fref{fig:msdss_alpha}, providing a
comparison between the two initial conditions analyzed in
\sref{sec:msd}, with the two corresponding sets of asymptotic regimes
given by equations \eqref{eq:numn2} and \eqref{eq:numn2ss}. The
scaling parameter values are set to $\alpha = 5/2$
(\fref{fig:msdss_alphagt2}), $2$ (\fref{fig:msdss_alphaeq2}), and
$3/2$ (\fref{fig:msdss_alphalt2}). The scattering parameter value
is set to $\epsilon = 1$ throughout, such that transitions between
scattering states are forbidden. As expected, the numerically computed
mean squared displacement obtained from the stationary initial
condition (magenta curves in figures
\ref{fig:msdss_alphagt2}-\ref{fig:msdss_alphalt2}) follow precisely
the analytic result \eqref{eq:n2ss} in all three regimes. 

We further note that, in the regime of normal diffusion ($\alpha=5/2$), both data
sets in \fref{fig:msdss_alphagt2} display consistent convergence to
the same asymptotic regime, given by the first lines of equations
\eqref{eq:numn2} and  \eqref{eq:numn2ss}, $\lim_{t\to\infty} \langle n^2
\rangle_t/t \simeq 2.597$. 
A similar result is observed in \fref{fig:msdss_alphaeq2}, where the
convergence of the two data sets to the common leading value of
equations  \eqref{eq:numn2} and \eqref{eq:numn2ss},
$\lim_{t\to\infty} \langle n^2 \rangle_t/t \log t = 4/5$, is
apparent. The effect of the differing subleading terms, the one in
\eqref{eq:numn2} negative, the other in \eqref{eq:numn2ss} positive
(the latter value is about one half the absolute value of the former
for $\delta = 4/3$), is, however, manifest. Finally, in the
superdiffusive regime, $\alpha =  3/2$, \fref{fig:msdss_alphalt2}
exhibits the two asymptotic regimes of the anomalously rescaled mean
squared displacement  given by equations \eqref{eq:numn2} and
\eqref{eq:numn2ss},  $\lim_{t\to\infty} \langle n^2 \rangle_t/t^{3 
  -\alpha} \simeq 0.582$ (all-scattering initial condition) or
$1.163$ (stationary initial condition), whose values
differ by a $1:2$ ratio for this value of the scaling parameter. 

\subsection{Perturbative regimes of the scattering
  parameter  \label{sec:numssvas}}  

\begin{figure}[p]
  \centering
  \begin{subfigure}[b]{0.49\textwidth}
  \includegraphics[width=\textwidth]
  {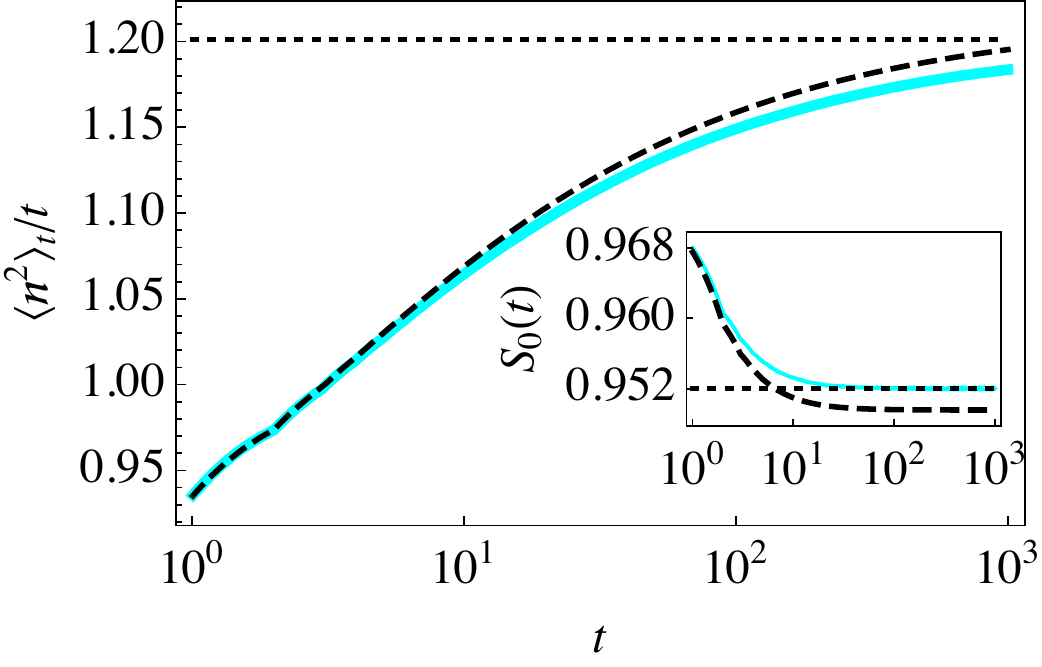}
  \caption{$\alpha = 5/2$, $\delta = 10^{-1}$  ($\epsilon \simeq
    3.4\times 10^{-2}$)} 
    \label{fig:msd_alphagt2_delta01}
  \end{subfigure}
  \hfill
  \begin{subfigure}[b]{0.49\textwidth}
    \includegraphics[width=\textwidth]
    {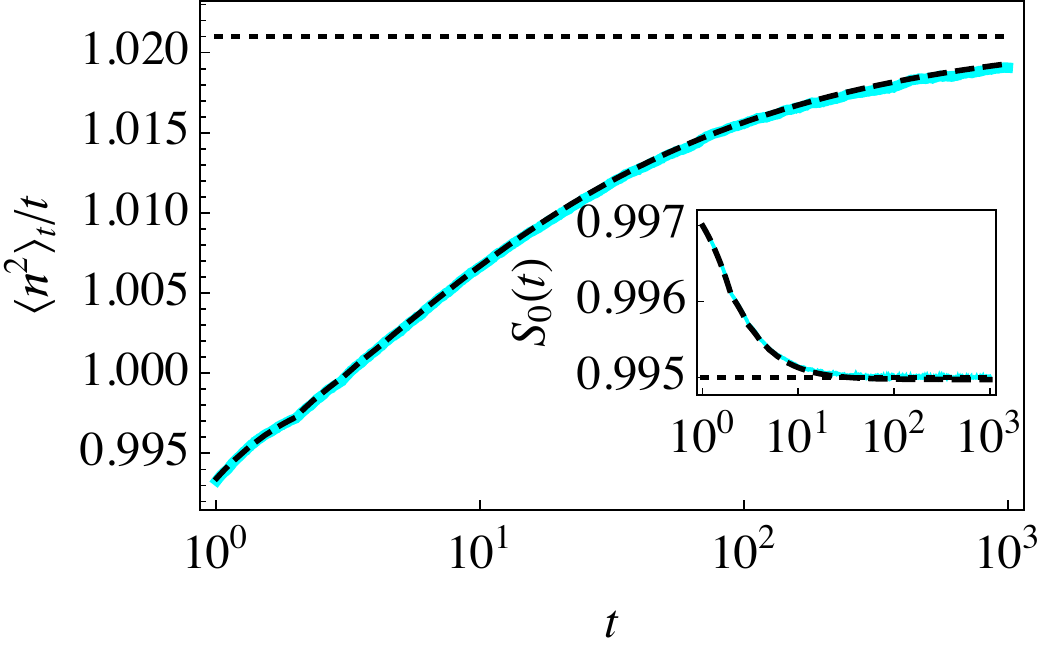}
    \caption{$\alpha = 5/2$, $\delta = 10^{-2}$ ($\epsilon \simeq
      3.25\times 10^{-3}$)} 
    \label{fig:msd_alphagt2_delta001}
  \end{subfigure}

  \begin{subfigure}[b]{0.49\textwidth}
    \includegraphics[width=\textwidth]
    {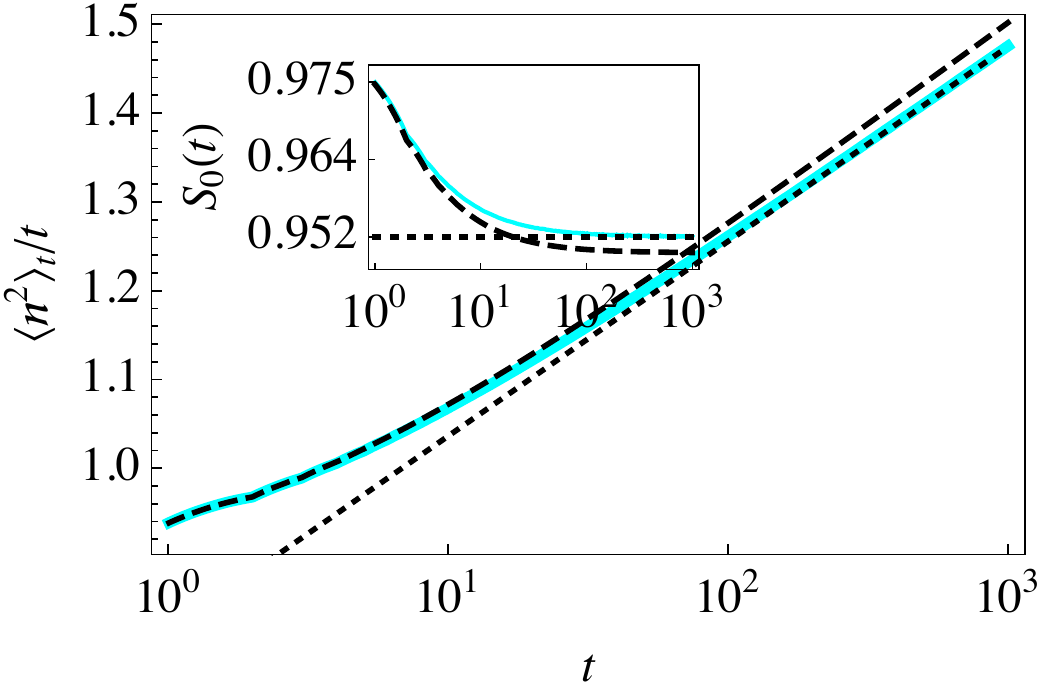}
  \caption{$\alpha = 2$, $\delta = 10^{-1}$ ($\epsilon \simeq
    2.63\times 10^{-2}$)} 
  \label{fig:msd_alphaeq2_delta01}
  \end{subfigure}
  \begin{subfigure}[b]{0.49\textwidth}
    \includegraphics[width=\textwidth]
    {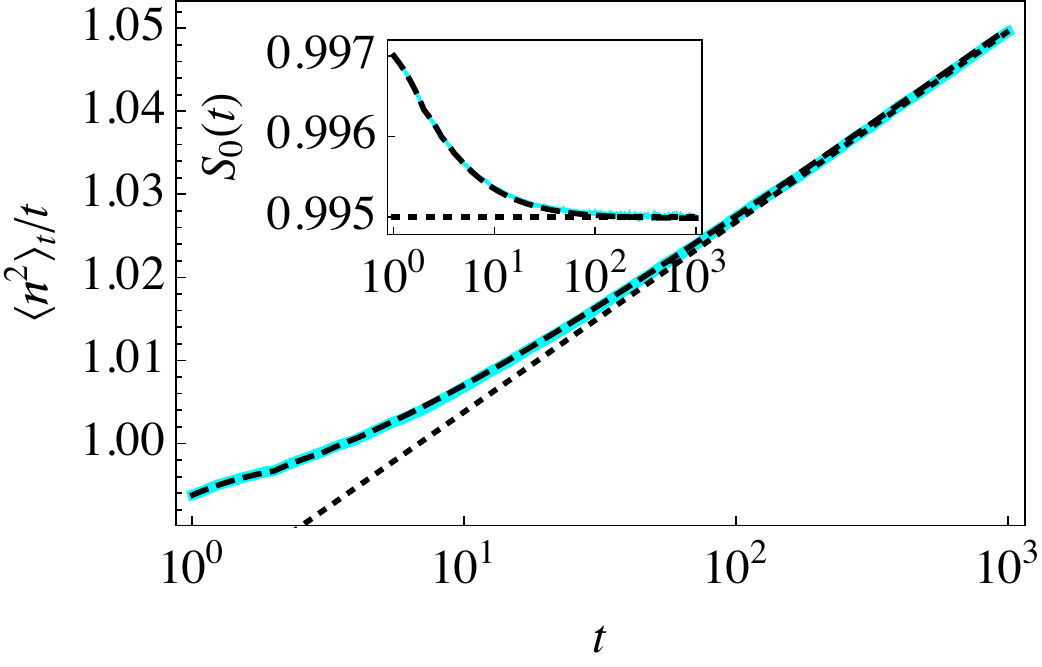}
    \caption{$\alpha = 2$, $\delta = 10^{-2}$ ($\epsilon \simeq
      2.51\times 10^{-3}$)} 
    \label{fig:msd_alphaeq2_delta001}
  \end{subfigure}
  
  \begin{subfigure}[b]{0.49\textwidth}
    \includegraphics[width=\textwidth]
    {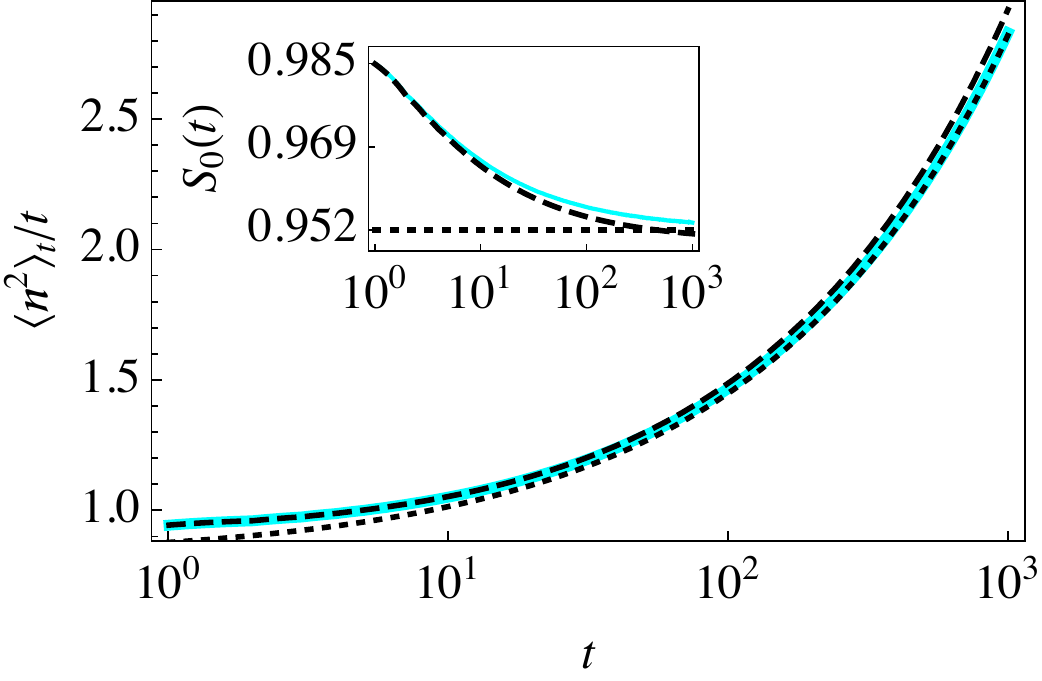}
    \caption{$\alpha = 3/2$, $\delta = 10^{-1}$ ($\epsilon \simeq
      1.54\times 10^{-2}$)} 
    \label{fig:msd_alphalt2_delta01}
  \end{subfigure}
  \begin{subfigure}[b]{0.49\textwidth}
    \includegraphics[width=\textwidth]
    {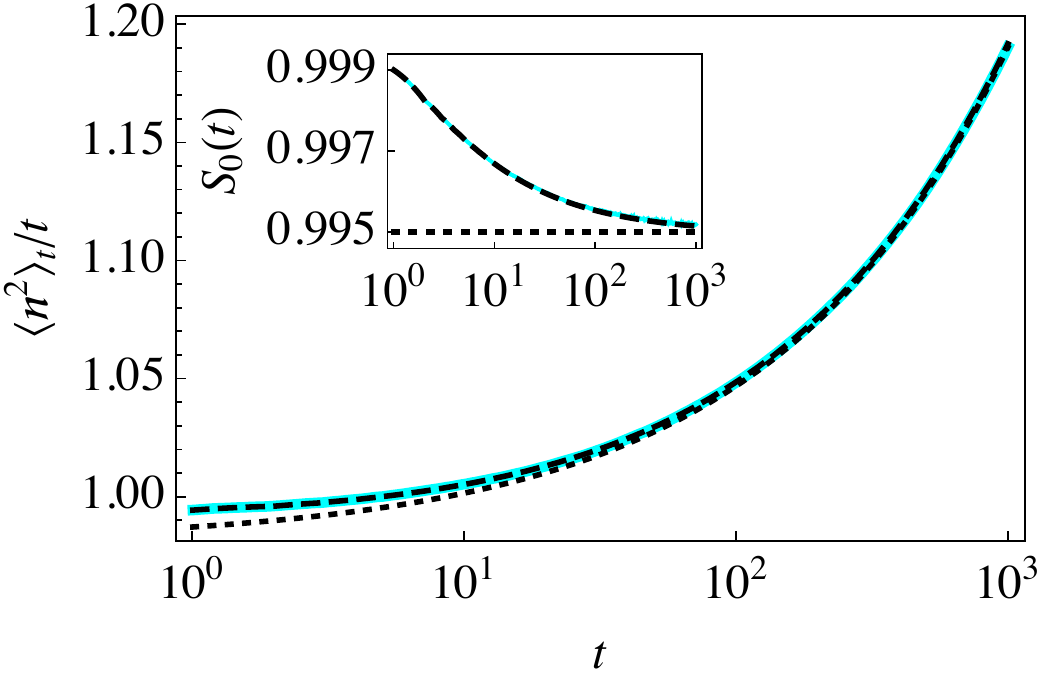}
    \caption{$\alpha = 3/2$, $\delta = 10^{-2}$ ($\epsilon \simeq
      1.47\times 10^{-3}$)} 
    \label{fig:msd_alphalt2_delta001}
  \end{subfigure}
  \caption{
    Numerical computations (cyan curves) of the normally rescaled mean
    squared displacement in the three different regimes of the scaling
    parameter $\alpha$ for the all-scattering initial condition. The
    scattering parameter takes values in or near the perturbative
    regime: (a), (c), (e) $\delta = 10^{-1}$,  and (b), (d), (f)
    $\delta = 10^{-2}$. The data are compared with the asymptotic
    solution \eqref{eq:numn2} (black dotted curves) and the
    $\Or(\epsilon)$ solution \eqref{eq:n2epsitime} (dashed curves). 
    Insets: time evolution of the fraction
    of particles in the scattering state with the stationary value
    \eqref{eq:S0rhok2} (dotted lines) and the the $\Or(\epsilon)$
    solution \eqref{eq:S0expepsi} (dashed curves).    
  }
  \label{fig:msd_alpha}
\end{figure}

The time evolution of the mean squared displacement in the
perturbative regime of the scattering parameter, $\delta \ll 1$, is
analyzed in \fref{fig:msd_alpha} for particles initially
distributed in the scattering state at the origin, where the rescaled
mean squared displacement is compared with  the $\Or(\epsilon)$
solution, equation  \eqref{eq:n2epsitime}. Excellent agreement
between the analytic and numerical results is observed throughout the
time range when the parameter is small enough ($\delta = 10^{-2}$
figures \ref{fig:msd_alphagt2_delta001},
\ref{fig:msd_alphaeq2_delta001},
\ref{fig:msd_alphalt2_delta001}). Numerical computations 
are also consistent with the asymptotic regime \eqref{eq:numn2} for
all cases.  

The scattering parameter value $\delta  = 10^{-1}$, shown  in 
figures \ref{fig:msd_alphagt2_delta01} ($\alpha = 5/2$),
\ref{fig:msd_alphaeq2_delta01} ($\alpha = 2$), and
\ref{fig:msd_alphalt2_delta01}  ($\alpha = 3/2$),  is on the one hand 
small enough that, for short times, the computed mean squared
displacement is barely distinguishable from the perturbative expansion
\eqref{eq:n2epsitime}. The effect of next order corrections is, on the
other hand, apparent for larger times ($ t \gtrsim  10^2$), beyond
which a convergence to the asymptotic result \eqref{eq:numn2} is
observed.

 We note that the crossover time \eqref{eq:timesupdiff} is
$t_\mathrm{c} \simeq 1.3 \times 10^4$ for $\delta = 10^{-1}$
(\fref{fig:msd_alphaeq2_delta01}) and much
larger yet for $\delta = 10^{-2}$. For $\alpha = 2$, subleading terms
in equation \eqref{eq:numn2} therefore dominate the logarithmically
divergent  term throughout the time range of measurements.
Similarly, in the superdiffusive case, $\alpha = 3/2$, the
respective crossover times \eqref{eq:timesupdiff} are $t_\mathrm{c}
\simeq 2 \times 10^2$ (\fref{fig:msd_alphalt2_delta01}) and
$t_\mathrm{c} \simeq 2.2 \times 10^{4}$ (\fref{fig:msd_alphalt2_delta001}),
such that subleading terms in equation \eqref{eq:numn2} remain
important throughout the time range of measurements in both cases. 

We conclude this section by pointing out that the time range of numerical
measurements such as presented in \fref{fig:msd_alpha} is limited by
the finite precision of the random number generator used to draw the
transition probabilities $\rho_k$ and sets an effective maximal scale
of free flights, $k_\mathrm{max}$. This observation is analogous to
the effect of machine-dependent limitations recently reported in
reference \cite{Radicchi:2014Underestimating} and is particularly
relevant to the regime of weak superdiffusion. The effective maximal
scale $k_\mathrm{max}$ induces a saturation of the logarithmic
growth of the second moment for times larger than $k_\mathrm{max}
\taub$. The process would thus become diffusive for times
sufficiently large. Although this effect can be pushed upward to
larger times by increasing the precision of the random number
generator, it cannot be eliminated altogether.  

\section{Conclusions}
\label{sec:con}

The inclusion of exponentially-distributed waiting times separating
the successive jump events of L\'evy walkers leads to a natural
distinction between propagating and scattering states, whose
respective concentrations evolve in time according to a set of
generalized master equations. In particular, the fraction of walkers 
in a scattering state obeys a linear delay differential equation with
a countable hierarchy of delays whose analytic solutions were
obtained. 

As opposed to classic methods based on the Fourier--Laplace transform
of an integral kernel and the use of Tauberian theorems
\cite{Geisel:1985p8023, Klafter:1987Stochastic}, the approach
presented in this paper is based solely on the solutions of such delay
differential equations. Our method thus yields a simple expression of the
mean squared displacement of L\'evy walkers in terms of the 
distribution of free paths and the time integral of the
fraction of particles in the scattering state.

Both exact and asymptotic expressions of the mean squared displacement
of walkers were obtained in regimes ranging from normal to
superdiffusive subballistic transport. In these regimes, the 
mechanism through which a walker can change directions between  
successive propagation phases plays an important role in determining
the values of the transport coefficients, whether normal or
anomalous. The transition through a scattering phase brings about a
description in terms of two parameters, the first specifying the
typical timescale of scattering events as opposed to the timescale of
propagation across an elementary cell, the second weighting the
probability of transitions between scattering and propagating states.

Relying on the stationary fractions of particles in scattering and
propagating states, a detailed derivation of the transport
coefficients, similar to that reported in
\cite{Cristadoro:2014transport}, was given, exhibiting the precise
effect of the two scattering parameters on these coefficients, as
well as the influence of the initial distribution of walkers
\cite{Zumofen:1993PowerSpectra}. Our formalism  also yields the exact
time evolution of the mean squared displacement, which was
investigated when particles are initially found in a scattering state. 

The comparison between the exact and asymptotic solutions is generally
not immediate, but is fairly straightforward in perturbative regimes
such that the likelihood of a transition from scattering to
propagating states is small.  A case in point -- though by far not the
only application of the present results -- is given by the 
infinite-horizon Lorentz gas \cite{Cristadoro:2014measuring}, for which the
scaling parameter of the distribution of free paths takes on the marginal
value $\alpha = 2$. As one might expect, measuring the logarithmic
divergence in time of the rescaled mean squared displacement is
typically hindered by dominant constant terms. This is particularly so
in the regime of narrow corridors, which yields a stochastic
description in terms of multistate L\'evy 
walks such that the scattering parameter is small, $\epsilon\ll1$
\cite{Cristadoro:2014Machta}. The parameter $\epsilon$ induces a
further separation between the time scales of the scattering and
propagating phases, $\taur\gg\taub$. To accurately model the
scattering phase is therefore of paramount importance to understanding
the dynamics of the infinite-horizon Lorentz gas in terms of a L\'evy
walk.   

The results obtained in this paper provide the framework to
transposing such results to a range of applications exhibiting
other regimes of transport, such as have been studied in the
context of optimal intermittent search strategies
\cite{Benichou:2011intermittent}. Our theoretical framework
brings about new perspectives to extend the study of intermittent
walks to power-law distributed ballistic phases, which may be relevant
to the increasing body of literature on optimal search strategies
\cite{Viswanathan:2011physics, Lewis:2013dispersal,
  Mendez:2013stochastic}.   

\ack
This work was partially supported by FIRB-Project No. RBFR08UH60 
(MIUR, Italy), by SEP-CONACYT Grant No. CB-101246 and DGAPA-UNAM
PAPIIT Grant No. IN117214 (Mexico), and by FRFC convention 2,4592.11
(Belgium). T.G. is financially supported by the (Belgian) FRS-FNRS.

\appendix

\section{Time-dependent fraction of particles in the scattering state \label{app:1}} 

To derive equation \eqref{eq:solS0}, we note that $a_{(1|k)} = \rho_k$
and, for $2\leq n \leq k$, 

\begin{equation}
  a_{(n|k)} = \frac{1}{n} \sum_{j = 1}^{k-n+1} \rho_{ j}
  a_{(n-1|k-j)}.
  \label{eq:akn_id1}
\end{equation}

\noindent By recursive application of this formula, we obtain

\begin{align}
  a_{(n|k)} &=  
              \frac{1}{n} \sum_{j_1 = 1}^{k-n+1} \rho_{ j_1}
              a_{(n-1|k-j_1)},
              \nonumber\\
            &=  
              \frac{1}{n(n-1)} \sum_{j_1 = 1}^{k-n+1} 
              \sum_{j_2 = 1}^{k-j_1-n+2} 
              \rho_{ j_1}  \rho_{ j_2}
              a_{(n-2|k-j_1-j_2)},
              \nonumber\\
            &=  
              \frac{1}{n!} 
              \sum_{j_1 = 1}^{k-n+1} \dots
              \sum_{j_{n-1} = 1}^{k-1-j_1-\dots-j_{n-2}} 
              \rho_{ j_i}  \dots \rho_{ j_{n-1}} \rho_{k - j_1 -\dots -j_{n-1}} ,
              \label{eq:akn_id2}
\end{align}

\noindent which is equivalent to equation \eqref{eq:akndef}.

Taking the derivative of equation \eqref{eq:solS0}, we verify equation
\eqref{eq:dtS0}:  

\begin{align}
  \dot \SZ(t) &= 
                - \epsilon\taur^{-1} \SZ(t)
                +  \sum_{k=1}^{\floor{t/\taub}} 
                e^{-\epsilon (t - k \taub)/\taur}  
                \sum_{n = 1}^{k} n\, a_{(n|k)}  \taur^{-n}  (t - k \taub)^{n-1} ,
                \nonumber\\
              &= 
                \taur^{-1}
                \sum_{k=1}^{\floor{t/\taub}} \rho_k \SZ(t - k \taub) - 
                \epsilon \taur^{-1}\SZ(t)    
                \nonumber\\
              & 
                + \sum_{k=1}^{\floor{t/\taub}} 
                e^{-\epsilon (t-k \taub)/\taur}  
                \Bigg[
                \sum_{n = 2}^{k} n\, a_{(n|k)} \taur^{-n} (t - k \taub )^{n-1}  
                \nonumber\\
              & 
                - \rho_k 
                \sum_{j=1}^{\floor{t/\taub}-k} e^{\epsilon j \taub/\taur}  
                \sum_{n = 1}^{j} a_{(n|j)} \taur^{-n-1} (t - k \taub - j \taub)^n 
                \Bigg].
\end{align}

\noindent We have to show that the last term on the RHS vanishes, which amounts
to proving the identity

\begin{align}
  \sum_{k=1}^{\floor{t/\taub}} 
  e^{\epsilon k \taub/\taur}  
  \Bigg[
  &
    \sum_{n = 2}^{k} n\, a_{(n|k)} \taur^{-n+1} (t - k \taub)^{n-1} 
    \nonumber\\
  & - \rho_k 
    \sum_{j=1}^{\floor{t/\taub}-k \taub} 
    e^{\epsilon j \taub/\taur}  \sum_{n = 1}^{j} 
    a_{(n|j)} \taur^{-n} (t - k \taub - j \taub)^n 
    \Bigg]
    = 0,
    \label{eq:id1}
\end{align}

\noindent or, after rearrangement,

\begin{align}
  \sum_{k=2}^{\floor{t/\taub}} 
  &
    e^{\epsilon k \taub/\taur}  
    \sum_{n = 1}^{k-1} (n+1) a_{(n+1|k)} \taur^{-n} (t - k \taub)^{n} 
    \nonumber\\
  &
    = \sum_{k=1}^{\floor{t/\taub}} 
    \rho_k 
    \sum_{j=1}^{\floor{t/\taub}-k} e^{\epsilon (k+j)\taub/\taur}  
    \sum_{n = 1}^{j} 
    a_{(n|j)} \taur^{-n} (t - k \taub - j \taub )^n.
    \label{eq:id2}
\end{align}

\noindent The second term in this expression, by successively changing the index
$j$  to $j-k$, then swapping the sums over $k$ and $j$ and exchanging
the indices $j$ and $k$,  transforms to

\begin{align}
  \sum_{k=1}^{\floor{t/\taub}} 
  & \rho_k 
    \sum_{j=1}^{\floor{t/\taub}-k} 
    e^{\epsilon (k+j)\taub/\taur}  
    \sum_{n = 1}^{j} 
    a_{(n|j)} \taur^{-n} (t - k \taub - j \taub)^n 
    \nonumber\\
  &
    = 
    \sum_{k=1}^{\floor{t/\taub}} 
    \rho_k 
    \sum_{j=k+1}^{\floor{t/\taub}} 
    e^{\epsilon j \taub/\taur}  
    \sum_{n = 1}^{j-k} 
    a_{(n|j-k)} \taur^{-n} (t - j \taub)^n ,
    \nonumber\\
  &
    = 
    \sum_{k=2}^{\floor{t/ \taub}} 
    e^{\epsilon k \taub/\taur}  
    \sum_{j=1}^{k-1} 
    \rho_j 
    \sum_{n = 1}^{k-j} 
    a_{(n|k-j)} \taur^{-n}
    (t - k \taub)^n .
\end{align}

\noindent Now swapping the sums over $j$
and $n$ in the last line and plugging this expression back into
equation \eqref{eq:id2}, we obtain, after factorization of the common
factors, the condition

\begin{equation}
  \sum_{k=2}^{\floor{t/\taub}} 
  e^{\epsilon k \taub/\taur}  
  \sum_{n = 1}^{k-1} 
  \Bigg[
  (n+1)  a_{(n+1|k)}
  - 
  \sum_{j=1}^{k-n} 
  \rho_j 
  a_{(n|k-j)}
  \Bigg] \taur^{-n} (t - k \taub)^{n}
  =  0,
  \label{eq:S0.id}
\end{equation}

\noindent 
which holds by identity \eqref{eq:akn_id1}, thus completing the proof
that $\SZ(t)$ as specified by equation \eqref{eq:solS0} is the solution to
equation \eqref{eq:dtS0} for the initial condition \eqref{eq:S0ic}.

\section*{References}


\providecommand{\newblock}{}

\end{document}